\documentclass[prb,twocolumn,showpacs,preprintnumbers,amsmath,amssymb]{revtex4}
\usepackage{graphicx}% Include figure files
\usepackage{dcolumn}% Align table columns on decimal point
\usepackage{bm}% bold math
\usepackage{float}
\usepackage{color}
\usepackage{setspace} 
\usepackage{arydshln}		%to draw dashed lines
\usepackage     [utf8]                  {inputenc}
\usepackage     [T1]                    {fontenc}
\usepackage                             {color}
\usepackage                             {amsmath}
\usepackage     [english]               {babel}
\usepackage     [usenames,dvipsnames]   {pstricks}
\usepackage{dcolumn}
\usepackage{bm}

\begin{document}

\input epsf
%%%%%%%%%%%%%%%%%%%%%%%%%%%%%%%%%%%%%%%%%%%%%%%%%%%%%%%%%%%%%%%%%%%%%%%%%%%%%%%%%%%%%%%%%%%%%%%%%
%%%%%%%%%%%%%%%%%%   NMR    COMMANDS   %%%%%%%%%%%%%%%%%%%%%%%%%%%%%%%%%%%%%%%%%%%%%%%%%%%%%%%%%%

%%%%%%%%%% Fields & Frequencies %%%%%%%%%%%%%
\newcommand{\Hz}{\ensuremath{H}_0}
\newcommand{\Hrf}{\ensuremath{H}_1}
\newcommand{\He}{\ensuremath{H}_e}
\newcommand{\Ome}[1][]{\ensuremath{\Omega_{e#1}}}
\newcommand{\om}[1][]{\ensuremath{\omega}}
\newcommand{\omo}[1][]{\ensuremath{\Omega_{0#1}}}
\newcommand{\omrf}[1][]{\ensuremath{\omega_{1#1}}}
%%%%%%%%%%% Spins %%%%%%%%%%%%%%%%%%%%%
\newcommand{\I}{\ensuremath{I'}}
\newcommand{\Idr}[2][]{\ensuremath{I^{#1\prime}_{#2}}}
\newcommand{\IIxyz}[5]{\ensuremath{\Idr{i #1}\Idr{j #2}#3\Idr{i #4}\Idr{j #5}}}
\newcommand{\IIpm}[5]{\ensuremath{\Idr[#1]{i}\Idr[#2]{j}#3\Idr[#4]{i}\Idr[#5]{j}}}
\newcommand{\rii}[2][2]{\ensuremath{r_{ij #2}^{#1}}}
%%%%%%%%% Agnles %%%%%%%%%%%%%%%%%%%%%%%%
\newcommand{\ca}{\ensuremath{\cos \alpha}}
\newcommand{\sa}{\ensuremath{\sin \alpha}}
\newcommand{\cb}{\ensuremath{\cos \beta}}
\newcommand{\sinb}{\ensuremath{\sin \beta}}
%%%%%%%%%%%%%% Hamiltonian %%%%%%%%%%%%%%
\newcommand{\Hdip}{\ensuremath{\mathcal{H}_{D}}}
\newcommand{\Ham}{\ensuremath{\mathcal}}
\newcommand{\Hdipo}{\ensuremath{\mathcal{H}_{D0}^l}}
\newcommand{\HdipOme}{\ensuremath{\mathcal{H}_{D1}^l}}
\newcommand{\HdipTOme}{\ensuremath{\mathcal{H}_{D2}^l}}
\newcommand{\HdipHOme}{\ensuremath{\mathcal{H}_{D3}^l}}
\newcommand{\HDu}[1]{\ensuremath{\mathcal{H}_{D#1}^u}}
\newcommand{\Hdipoo}{\ensuremath{\HDu{0}}}
%%%%%%%%%%%%%% Other %%%%%%%%%%%%%%
\newcommand{\VEC}{\boldsymbol}
\newcommand{\Rfactor}{\ensuremath{J_{ij}}}
\newcommand{\T}{\ensuremath{\tau}}

%%%% Supplement commands

%Abreviation for often used symbols

%The RF field. V stands for vector
\newcommand{\BrfV}{
\ensuremath{\VEC{H}_1}
}
\newcommand{\Brf}{
\ensuremath{H_1}
}
%The static field. V stands for vector.
\newcommand{\BzV}{
\ensuremath{\VEC{H}_0}
}
\newcommand{\Bz}{
\ensuremath{H_0}
}

%An arbitrary magnetic fields
\newcommand{\BV}{
\ensuremath{\VEC{H}}
}
%The effective magnetic field in the rotating frame. V stands for vector
\newcommand{\BeV}{
\ensuremath{\VEC{H}_e}
}
\newcommand{\Be}{
\ensuremath{H_e}
}

%The fields in frequency
%The frequency amplitudes can have an additionnal index. Just add [-index-] to the command. Eg : \Omz[2]=\Omega_{02}

%The Larmor frequency. V stands for vector.
\newcommand{\OmzV}[1][]{
\ensuremath{\VEC{\Omega}_{0#1}}
}

\newcommand{\Omz}[1][]{
\ensuremath{\Omega_{0#1}}
}

%The RF frequency. V stands for vector.
\newcommand{\OmrfV}{
\ensuremath{\VEC{\omega}_1}
}

\newcommand{\Omrf}[1][]{
\ensuremath{\omega_{1#1}}
}

%The RF fields radio frequency. V stands for vector.
\newcommand{\OmV}{
\ensuremath{\VEC{\omega}}
}
\newcommand{\Om}[1][]{
\ensuremath{\omega_{#1}}
}

%The effective Larmor frequency. V stands for vector.
\newcommand{\OmeV}[1][]{
\ensuremath{\VEC{\Omega}_{e#1}}
}

%============================================

%An arbitrary Hamiltonian

%The names of the recoupled dipolar Hamiltonians. 
%Hldip for like spins
\newcommand{\Hldip}[1][]{
{\ensuremath{\mathcal{H}^{\text{l}}_{D#1}}}
}
%Hudip for unlike spins
\newcommand{\Hudip}[1][]{
{\ensuremath{\mathcal{H}^{\text{u}}_{D#1}}}
}

%===============================================

%Function for vectors notation and so on

%Vector notation

%=============================================

% Upright e for exponentialfunktion
\newcommand{\e}
{\textrm{e}}

%Unit internuclear distance vector

\newcommand{\n}{
\ensuremath{n}
}

%The variable name of the dipole-dipole Hamiltonian prefactor
\newcommand{\J}{
J
}

%The variable name of the torque (not M, as it is reserved for the total magnetization)

\newcommand{\Tor}{
\ensuremath{\mathcal{M}}
}

%A spin in the effective stationnary Larmor coordinates
%V stands for vector.
\newcommand{\IV}{
\ensuremath{\VEC{I}'}
}

%The total magnetization effective stationnary Larmor coordinates 
%V stands for vector.
\newcommand{\MV}{
\ensuremath{\VEC{M}'}
}
\newcommand{\M}{
\ensuremath{M'}
}

%The spin sequence I_i,xI__j,y+I_i,yI_j,x  with -x- -y- -+- -y- -x- the five arguments

%The same as before but withtout primes
\newcommand{\sIIxyz}[5]{
\ensuremath{I_{i,#1}I_{j,#2}#3I_{i,#4}I_{j,#5}}
}

%The arccot
\newcommand{\arccot}{
\ensuremath{\hbox{arccot}}
}

%Transformation matrices
\newcommand{\U}[2]{
\ensuremath{{U}^{#1}_{#2,\mu\nu}}
}

%Transformation matrices
\newcommand{\Cij}{
\ensuremath{C_{\mu\nu}}
}
%Transformation matrices
\newcommand{\A}{
\ensuremath{A_{\mu\nu}}
}
%Transformation matrices
\newcommand{\Pij}{
\ensuremath{P_{\mu\nu}}
}
%Transformation matrices
\newcommand{\Q}{
\ensuremath{Q_{\mu\nu}}
}
%%%%%%%%%%%%%========%%%%%%%%%%%%%%5

%Recoupled Hamiltonians references
\newcommand{\Reslike}{\eqref{eq:Om=Ome}-\eqref{eq:2Om=-Ome} }
\newcommand{\ResunlikeO}{\eqref{eq:Om=Omei}-\eqref{eq:2Om=-Omei} }
\newcommand{\ResunlikeT}{\eqref{eq:Om=Ome1+Ome2}-\eqref{eq:2Om=Ome1-Ome2} }

\newcommand{\Hlike}{\eqref{eq:Hldip(om=Ome)}-\eqref{eq:Hldip(2om=Ome)} }
\newcommand{\HunlikeO}{\eqref{eq:Hudip(om=Ome)}-\eqref{eq:Hudip(2om=Ome)} }
\newcommand{\HunlikeT}{\eqref{eq:Hudip(om=(Ome1+-Ome2))}-\eqref{eq:Hudip(2om=-(Ome1+-Ome2))} }

%%%%%%%%%%%%%%%%%%   NMR    COMMANDS   %%%%%%%%%%%%%%%%%%%%%%%%%%%%%%%%%%%%%%%%%%%%%%%%%%%%%%%%%%
%%%%%%%%%%%%%%%%%%%%%%%%%%%%%%%%%%%%%%%%%%%%%%%%%%%%%%%%%%%%%%%%%%%%%%%%%%%%%%%%%%%%%%%%%%%%%%%%%

%\draft

%\twocolumn[\hsize\textwidth\columnwidth\hsize\csname %
%@twocolumnfalse\endcsname

\title{Recoupling the non-secular part of nuclear spin-spin interaction in solids}

\author{Chahan M. Kropf}
\email{C.Kropf@thphys.uni-heidelberg.de}
\affiliation{Institute for Theoretical
Physics, University of Heidelberg, Philosophenweg 19, 69120
Heidelberg, Germany}
\author{Boris V. Fine}
\email{B.Fine@thphys.uni-heidelberg.de}
\affiliation{Institute for Theoretical
Physics, University of Heidelberg, Philosophenweg 19, 69120
Heidelberg, Germany}

\date{February 6, 2012}

\begin{abstract}

NMR spin-spin relaxation in solids in strong magnetic fields is normally described only with the help of the secular part of the full spin-spin interaction Hamiltonian. This approximation is associated with the averaging of the spin-spin interaction over the fast motion of spins under the combined action of the static and the radio-frequency (rf) fields. Here we report a set of conditions (recoupling resonances) when the averaging over the above fast motion preserves some of the non-secular terms entering the full interaction Hamiltonian. These conditions relate the value of the static magnetic field with the frequency and the amplitude of the rf field. When the above conditions are satisfied, the effective recoupled spin-spin interaction Hamiltonian has an unconventional form with tunable parameters. These tunable Hamiltonian offers interesting possibilities to manipulate nuclear spins in solids and can shed new light on the fundamental properties of the nuclear spin-spin relaxation phenomenon. 
\end{abstract}

%\vskip 2cm\centerline{PLEASE DO NOT DISTRIBUTE.}
%\vskip 2cm

\maketitle

%\narrowtext
%\pagebreak
\section{Introduction}
The experimental technique of nuclear magnetic resonance (NMR) has long established itself as a versatile tool for microscopic investigations of matter. One major source of microscopic information in NMR is the spin-spin relaxation. In solids, the description of NMR spin-spin relaxation requires averaging the full Hamiltonian of nuclear spin-spin interaction over the fast spin rotations induced by strong external magnetic fields.  The present paper predicts theoretically a set of resonant conditions for solid-state NMR experiments leading to an unconventional form of the averaged interaction Hamiltonians.

We consider nuclear spin-spin interactions in the presence of a static magnetic field and a continuously applied  {\it rotating}  radio-frequency (rf) field in the regime, when the two fields have comparable values, while both of them are still much larger than the local fields with which nuclear spins affect each other.  Such a setting is realizable experimentally. However, it implies the use of smaller-than-typical static fields and, therefore, a certain sacrifice in terms of the signal intensity.  In a typical NMR experiment, the rf-field may be large in comparison with the local fields, but it is still much smaller than the static field. 
This ``typical'' setting for strong continuously applied rf fields was originally considered by Redfield\cite{Redfield-55} and subsequently discussed in NMR textbooks\cite{Abragam-61,Slichter-90}. In such a case, the averaged interaction Hamiltonian is obtained from the full Hamiltonian with the help of a two-step truncation procedure\cite{Redfield-55}. The first step is to eliminate all interaction terms that do not commute with the Zeeman Hamiltonian for the static magnetic field\cite{VanVleck-48}. The second step is to make the transformation into the rotating reference frame where the rf field appears static, and then further truncate the interaction Hamiltonian by eliminating all the terms that do not commute with the Zeeman Hamiltonian for the effective magnetic field in the rotating frame. 

When the static and the rf fields are comparable to each other, the validity of the above two-step truncation becomes questionable, because the spin motions caused by both fields fall on the same timescale. The full interaction Hamiltonian then needs to be averaged over the above two motions simultaneously rather than in two independent steps. In this paper, we show that, in the case of a purely rotating rf field (as opposed to the case of a linearly-polarized rf field), such a single step averaging can be done rigorously without requiring the rf field to be smaller than the static field. We find that such a single-step procedure generically reproduces the results of Redfield's two-step truncation, but the reason for this agreement lies not in the perturbation theory but in the ergodic character of the single-spin dynamics driven by the static and the rf fields. We also find that in certain special cases, which we call ``recoupling resonances'', the averaged spin-spin interaction Hamiltonian includes other terms in addition to the outcome of Redfield's truncation procedure.  In the case of single spin species, the recoupling resonances appear when the rf field frequency is matched with certain multiples or simple fractions of the Larmor frequency corresponding to the effective magnetic field in the rotating reference frame. In the case of two different spin species, the sums and the differences of the two effective Larmor frequencies also appear in some of the recoupling resonance conditions.  
The time-averaged interaction Hamiltonians associated with the above recoupling resonances are non-secular, in the sense that they do not conserve the total Zeeman energy associated with the external fields in either the laboratory or the rotating reference frame. These Hamiltonians may have a range of applications, such as, e.g., nuclear cross-polarization, or molecular structure determination. 

It is worth noting that the term ``recoupling'' is extensively used\cite{Oas-88,Griffiths-93,Scholz-07,Bayro-09} in the context of the magic angle spinning (MAS) technique of NMR, where it implies the recovery of a part of the standard truncated Hamiltonian, after that Hamiltonian was suppressed by the spinning of the sample. The MAS recoupling schemes match the frequency of the spin nutation\cite{Oas-88} or the frequency of the rf pulses\cite{Griffiths-93,Scholz-07,Bayro-09} with the multiples of the sample spinning frequency. In the present paper, the term ``recoupling'' is used at a more basic level: it implies the recovery of a part of the {\it full} spin-spin interaction Hamiltonian lost in the standard truncation procedure.
 The recoupling resonances described below have nothing to do with sample spinning. However, they are best observable and, probably, most useful in combination with MAS.   

In a broader context, our results could have been obtained in the framework of the averaged Hamiltonian theory\cite{Haeberlen-68}, but to the best of our knowledge it has never been attempted, perhaps because of the non-perturbative character of the problem.  The role of the non-secular terms was, however, discussed in the context of the phenomenon of nuclear spin superradiance\cite{Yukalov-95,Yukalov-96}.

The rest of this paper is organized as follows. We give the general formulation of the problem in Section~\ref{general}, introduce our single-step truncation procedure in Section~\ref{truncation}, present the results of this truncation with the identification of the recoupling resonances in Section~\ref{results}, and then discuss a possible experimental test and possible applications of the recoupling resonances in Section~\ref{discussion}. 

\section{General formulation}
\label{general}

We consider a lattice of nuclear spins in a static magnetic field pointing along the 
$z$-axis, $\VEC{H}_0 \equiv (0,0,H_0)$, irradiated by a rf field rotating in the $x$-$y$ plane with frequency $\om$,  \mbox{$\VEC{H}_1(t) \equiv (H_1 \hbox{cos} \, \omega t, H_1 \hbox{sin} \, \omega t,0) $}. We limit ourselves to the case of the magnetic dipole spin-spin interaction, which is, typically, dominant in solids. The full Hamiltonian \cite{Abragam-61,Slichter-90} is 
\begin{equation}
\Ham{H}= \Ham{H}_z+\Ham{H}_{\hbox{\small rf}}+\Hdip
\label{Htot}
\end{equation}
where 
\begin{equation}
\Ham{H}_z=- H_0 \sum_{i}^N \gamma_i I_{iz},
\label{Hz}
\end{equation}
\begin{equation}
\Ham{H}_{\hbox{\small rf}}= -\VEC{H}_1(t) \cdot \sum_{i}^N  \gamma_i \VEC{I}_i,
\label{Hrf}
\end{equation}
and 
\begin{equation}
\Hdip =\sum_{i<j}^N J_{ij}\; [\VEC{I}_i \cdot \VEC{I}_j-\frac{3\cdot(\VEC{I}_i \cdot\VEC{r}_{ij})(\VEC{I}_j \cdot\VEC{r}_{ij})}{r_{ij}^2}].\label{eq:Hdip}
\end{equation}
Here $J_{ij} = \frac{\gamma_i\gamma_j \hbar^2}{r_{ij}^3}$ are the coupling constants, $\VEC{I}_i \equiv (I_{ix}, I_{iy}, I_{iz})$ are the operators of the nuclear spin of the $i$th lattice site, $\gamma_i$ is the gyromagnetic ratio of that spin, $\VEC{r}_{ij} \equiv (r_{ij,x},r_{ij,y},r_{ij,z})$ are the displacement vectors between two lattice sites $i$ and $j$, and $N$ is the number of spins. The characteristic time of a single spin motion induced by the Hamiltonian $\Hdip$ can be estimated as $T_2=\hbar\left( \sum_j J_{ij}^2 \right)^{-1/2}$. The variable $T_2$ is also to be referred to as the transverse relaxation time.  The timescale of nuclear spin relaxation due to the coupling to electrons or phonons is to be denoted as $T_1$ (longitudinal relaxation time). We further introduce the Larmor frequencies $\Omega_{0i} = \gamma_i H_0$ and the nutation frequencies $\omega_{1i} = \gamma_i H_1$. In the following, we drop lattice index $i$ in the gyromagnetic ratio and use the notations
$\gamma \equiv \gamma_i$, $\Omega_0 = \gamma H_0$, $\omega_1 = \gamma H_1$, etc., when not dealing with the interactions of non-equivalent (unlike) spins explicitly.

This paper is limited to the regime $\Omega_0, \omega_1 \gg 1/T_2 \gg 1/T_1$. The first of these inequalities implies the separation of the timescales between the fast motion due to the external fields and the slow motion due to the spin-spin interaction. The second inequality allows us to consider the nuclear spins as isolated from the environment. As mentioned in the introduction, the theoretical analysis in the literature\cite{Redfield-55,Abragam-61,Slichter-90} was limited so far by the additional inequality $H_1 \ll H_0$ (i.e. $\omega_1 \ll \Omega_0$). Our main focus is on the regime $H_1 \sim H_0$. However, also, for a range of situations satisfying the inequality $H_1 \ll H_0$, the recoupling resonances obtained below will imply important observable effects.

\section{Truncation procedure}
\label{truncation}

We calculate the effective spin-spin interaction Hamiltonian by averaging the full interaction Hamiltonian $\Hdip$ over the fast spin motion induced by the terms $\Ham{H}_z+\Ham{H}_{\hbox{\small rf}}$. In the Heisenberg representation, this fast motion is described by the following equation for each of the spins~:
\begin{equation}
 { d \VEC{I}_i \over dt}=\VEC{I}_i\times\gamma\VEC{H}(t), 
\label{eq:Spin-eom}
\end{equation}
where $\VEC{H}(t) = \VEC{H}_0 + \VEC{H}_1(t) $. The above equation is linear in terms of 
spin operators $(I_{ix}, I_{iy}, I_{iz})$, and, therefore, its predictions are essentially the same as those for the classical spins, in which case  $(I_{ix}, I_{iy}, I_{iz})$ would be simply three numbers. Despite the linearity of Eq.(\ref{eq:Spin-eom}), it is not solvable analytically either classically or quantum mechanically for an arbitrary time-dependent $\VEC{H}(t)$. However, for the specific time dependence considered in this work, the solution is facilitated by the transformation from the laboratory reference frame into the reference frame rotating around the $z$-axis with frequency $\omega$. The equations of motion in the rotating frame  have the same form as Eq.(\ref{eq:Spin-eom}), but with time-independent effective magnetic field $\VEC{H}_e = (H_1, 0, H_0 + \omega/\gamma)$. (The $x$-axis in the rotating frame coincides with the direction of the rf field.) The corresponding Larmor frequency is
\begin{equation}
	\Ome = \gamma \He= \hbox{sign} (\gamma) \  \sqrt{\omrf^2+(\omo+\om)^2}.
\label{eq:Ome}
\end{equation}
Thus, the motion of a spin in the laboratory frame can be characterized as the Larmor precession with frequency $\Ome$ around the direction of $\VEC{H}_e$, which itself rotates with frequency $\omega$ around the $z$-axis.

Our formal procedure for averaging $\Hdip$ consists of the transformation into the reference frame that follows the fast spin rotation induced by Eq.(\ref{eq:Spin-eom}). In that reference frame, each spin would be motionless, if the full Hamiltonian included only $\Ham{H}_z+\Ham{H}_{\hbox{\small rf}}$. Such a transformation is time-dependent. As a result, the
Hamiltonian $\Hdip$ expressed in the new spin coordinates acquires some quickly oscillating interaction coefficients, which we then average out. 

The above procedure may, at first sight, appear equivalent to the one introduced by Redfield\cite{Redfield-55,Abragam-61,Slichter-90}. The difference, however, is that Redfield did not define the entire transformation and then average the resulting Hamiltonian over time but rather used the Zeeman-energy-conservation criterion of Van Vleck\cite{VanVleck-48} to average separately the Hamiltonians obtained at the intermediate steps of this transformation. Such an independent averaging of the intermediate Hamiltonians is justifiable by the perturbation theory arguments~\cite{Redfield-55} only when $\Omega_0 \gg \omega_1$. It is shown below, that in the case of $\Omega_0 \sim \omega_1$, the spin motions associated with the intermediate transformation steps may be correlated, and thus the averaging over them cannot be done independently. 

\begin{figure} \setlength{\unitlength}{0.1cm} 
%=======================================================================

\begin{picture}(80, 50)
{ 
\put(0,0){ \epsfxsize= 3.3cm \epsfbox{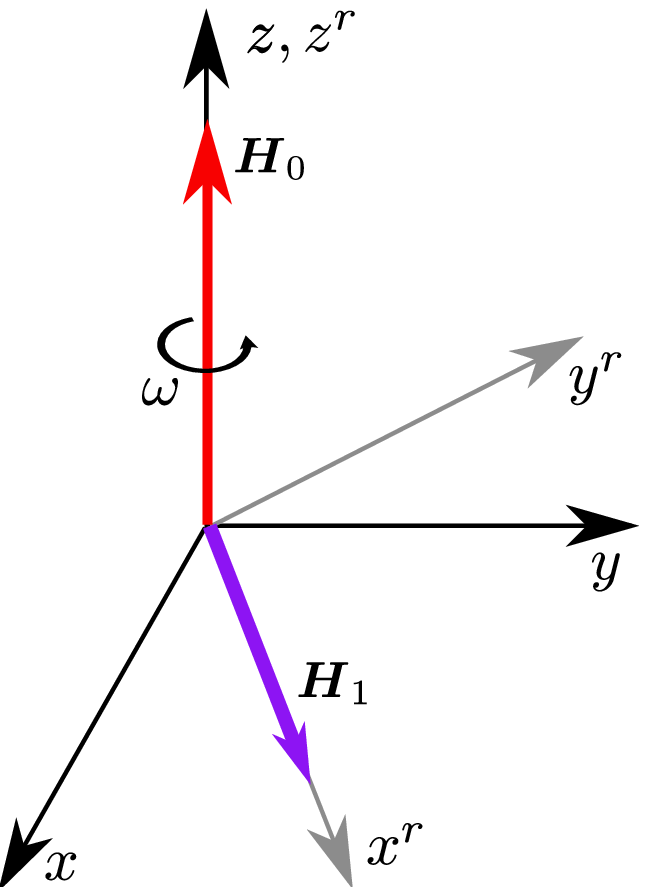} }
\put(0,43){(a)}
\put(40,1){ \epsfxsize= 3.5cm \epsfbox{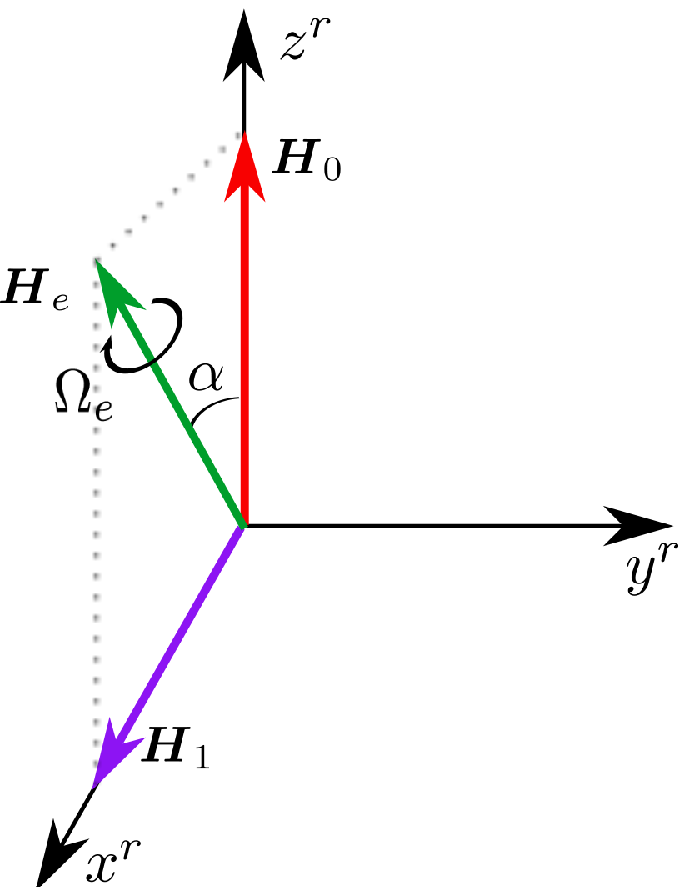} }
\put(40,43){(b)}
}
\end{picture} 
%============== 
\caption{(a) Illustration of the transformation from the laboratory reference frame $(x,y,z)$ to the rotating reference frame $(x^r,y^r,z^r)$ --- Eq.(\ref{L2R}). (b) Illustration for the transformation from the rotating reference frame $(x^r,y^r,z^r)$ to the tilted-rotating and then double-rotating reference frames --- Eqs.(\ref{R2TR},\ref{TR2DR}).}
\label{fig:Frames}
\end{figure}

The overall transformation into the reference frame that follows the motion induced by Eq.(\ref{eq:Spin-eom}) can be decomposed into three simple transformations visualized in Fig.\ref{fig:Frames}. The first transformation is from the laboratory frame to the reference frame rotating with the rf field:
\begin{equation}
U_{\om}=\begin{pmatrix} \cos(\om t) & \sin (\om t) & 0 \\ -\sin(\om t) & \cos(\om t) & 0 \\ 0 & 0 & 1\end{pmatrix} .
\label{L2R}
\end{equation}
The second transformation is from the rotating to the ``tilted-rotating'' reference frame, where the $z$-axis coincides with the direction of  $\VEC{H}_e$:
\begin{equation}
U_{\alpha}=\begin{pmatrix} \cos \alpha & 0 & -\sin \alpha \\ 0 & 1 & 0\\ \sin \alpha  & 0 & \cos \alpha \end{pmatrix},
\label{R2TR}
\end{equation}
with 
\begin{equation}
\alpha = \hbox{arccot}\left[ \frac{\Omega_0 + \omega}{\omega_1} \right],
\label{alpha}
\end{equation}
admitting values in the range $[0, \pi]$.

Finally, the third transformation is into the ``double-rotating'' reference frame precessing around $\VEC{\He}$ with frequency $-\Ome$: 
\begin{equation}
U_{\Ome}=\begin{pmatrix} \cos(\Ome t) & - \sin(\Ome t) & 0 \\ \sin(\Ome t) & \cos(\Ome t) & 0 \\ 0 & 0 & 1\end{pmatrix} .
\label{TR2DR}
\end{equation}
The effective magnetic field in this reference frame is equal to zero. The explicit implementation of transformation (\ref{TR2DR}) before performing the time-averaging of $\Hdip$ formally discriminates our treatment from that of Redfield\cite{Redfield-55,Abragam-61,Slichter-90}.

%%%%%%%%%%%%%%%%%%%%%%%%%%%%%%%%%%%%%%%%%%%%%%%%%%%%%%%%%%%%%%%%%%%
%%%%%%%%%%%%%%%%%%%%%%%%%%%%%%%%%%%%%%%%%%%%%%%%%%%%%%%%%%%%%%%%%%%

We denote the axes in the double-rotated reference frame as $\{x',y',z'\}$, and the corresponding spin projections as $\VEC{I}'_i \equiv (I'_{ix},I'_{iy},I'_{iz})$.
The spin coordinates in the laboratory reference frame can now be expressed as
\begin{equation} 
 \VEC{I}_i = U_{\om}^{-1}U_{\alpha}^{-1}U_{\Ome}^{-1} \; \VEC{I}_i'.
 \label{eq:IIp}
\end{equation}
The substitution of Eq.(\ref{eq:IIp}) into Eq.\eqref{eq:Hdip} allows us to reexpress the interaction Hamiltonian $\Hdip$ in terms of $\VEC{I}_i'$ and then average the interaction coefficients in the resulting expression.

The full time-dependent form of $\Hdip$ in the double-rotated reference frame is rather long. It can be found in Ref.\cite{supplement} together with the implementation of the time-averaging procedure.  Below we exemplify this procedure by one typical term and then, in the next section, present only the resulting time-averaged expressions for $\Hdip$. 

The term we have chosen as an example, originates from the calculation for equivalent (like) spin species discussed in the next section. It has form 
\begin{equation}
 a_{ij}(t) \; c(\VEC{r}_{ij}) \; I_{ix}'I_{jx}',  
\label{eq:Hterm}
\end{equation}
where $c(\VEC{r}_{ij})$ is some function of the relative displacement between the two nuclei, and
\begin{equation}
a_{ij}(t)= \cos^2 (\om t) \cos^2 (\Ome t).  
\label{eq:A12Average}
\end{equation}
The time-averaged value of the above coefficient is calculated as follows 
\begin{equation}
 \langle a_{ij} \rangle = \lim_{\T \to \infty} \frac{1}{2\T}\int_{-\T}^{\T} a_{ij}(t)dt
 = \begin{cases}
 \frac{1}{4}&\text{if } \om \neq \pm\Ome ,\\
\frac{3}{8} &\text{if } \om=\pm\Ome . 
\end{cases}
\label{eq:Aver}
\end{equation}
In the above case, $\langle a_{ij} \rangle = \frac{1}{4}$ is what one obtains by the standard two-step truncation procedure outlined in the introduction, while, in the case  of $\om=\pm\Ome$, there is an extra contribution associated with the fact that 
$\langle \cos^2 (\om t) \cos^2 (\Ome t) \rangle \neq \langle \cos^2 (\om t) \rangle \langle \cos^2 (\Ome t) \rangle$.

The question then arises, what happens, if, in an experiment, the actual values of $H_0$, $H_1$ and $\omega$ are such that one of the
recoupling resonance conditions is satisfied approximately but not exactly. 
In such a case, the Hamiltonian-averaging procedure needs to be modified. Below, we illustrate this modification for the resonant condition $\om = - {1 \over 2} \Ome$ but the expression for the resulting correction to the averaged Hamiltonian is applicable to all other recoupling resonances.  

In the chosen example, the mismatch of the resonant condition can be parameterized with the help of variable $\Delta \Omega_e$ defined by equation $\om = - {1 \over 2} (\Ome + \Delta \Ome) $. Now we modify the last step  of the transformation to the double-rotated reference frame [Eq.(\ref{TR2DR})] by changing the value of the rotation frequency from $\Ome$ given by Eq.(\ref{eq:Ome}) to $\Ome + \Delta \Ome $. In the modified double-rotated reference frame, the averaged interaction Hamiltonian obtained in Section~\ref{like} for the recoupling resonance $\om = - {1 \over 2} \Ome$ will be exactly valid, but, in addition, since $\Ome + \Delta \Ome$ is different from $\Ome$, the effective magnetic field will be different from zero, which then gives rise to the following additional term:
\begin{equation}
\Delta {\cal H} = \Delta \Ome  \sum_i I'_{iz}.
\label{DeltaH}
\end{equation}

When $\Delta \Ome \gg 1/T_2$, such a term suppresses the effect of the resonant non-secular terms given in the next section. This is, of course, expected for a significant departure from the recoupling resonance conditions. When $\Delta \Ome \sim 1/T_2$, the above term still suppresses the effect of the non-secular terms but only partially, and this partial suppression is of the same order of magnitude as the suppression caused by the secular interaction terms (to be explained in Section~\ref{like}). In such a case, the term (\ref{DeltaH}) should simply be included in the averaged Hamiltonian together with the secular terms and the resonant non-secular terms. 

The above considerations imply, that, in a generic case, when the strength of the secular terms can be estimated as $\hbar/T_2$, the recoupling resonance conditions cannot be resolved experimentally with accuracy much better than $\pm 1/T_2$. However, if the secular terms are suppressed by other means, then the experimental resolution will be limited by the strength of the recoupled terms themselves. Thus, if the strength of the recoupled terms is much smaller than $\hbar/T_2$, then the recoupling resonances can be resolvable with proportionally better accuracy.

%%%%%%%%%%%%%%%%%%%%%%%%%%%%%%%%%%%%%%%%%%%%%%%%%%%%%%%%%%%%%%%%%%%
\begin{figure}[t] \setlength{\unitlength}{0.1cm} 
%=======================================================================
\begin{picture}(80, 72)
{ 
\put(0,36){ \epsfxsize= 3.5cm \epsfbox{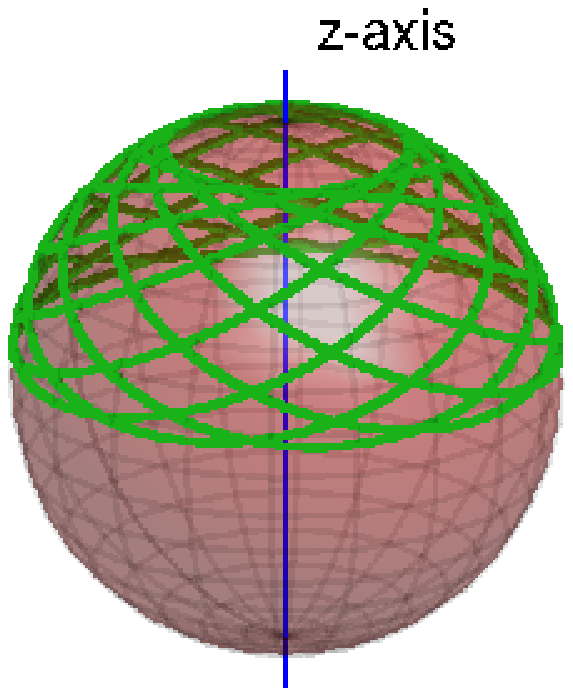} }
\put(0,63){{\large (a)}}
\put(40,39){ \epsfxsize= 3.5cm \epsfbox{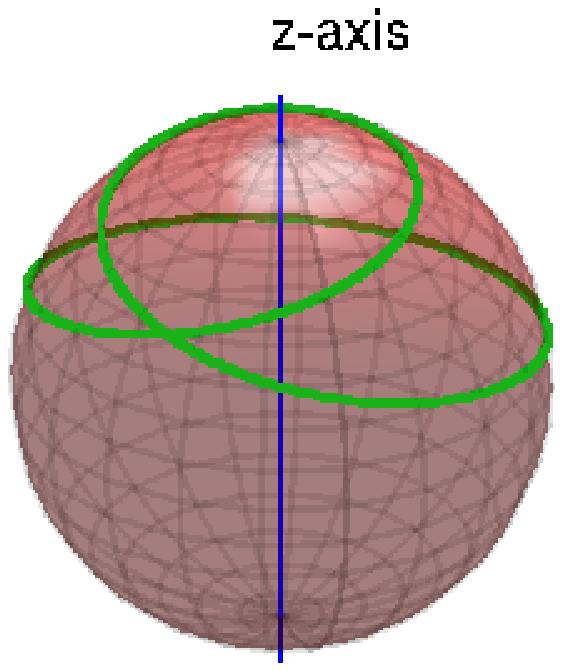} }
\put(40,63){{\large (b)}}
\put(0,0){ \epsfxsize= 3.5cm \epsfbox{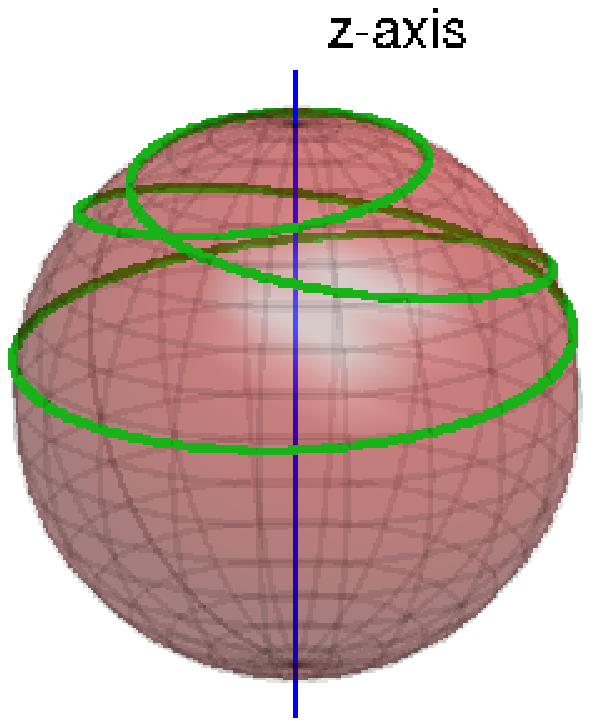} }
\put(0,29){{\large (c)}}
\put(40,1){ \epsfxsize= 3.5cm \epsfbox{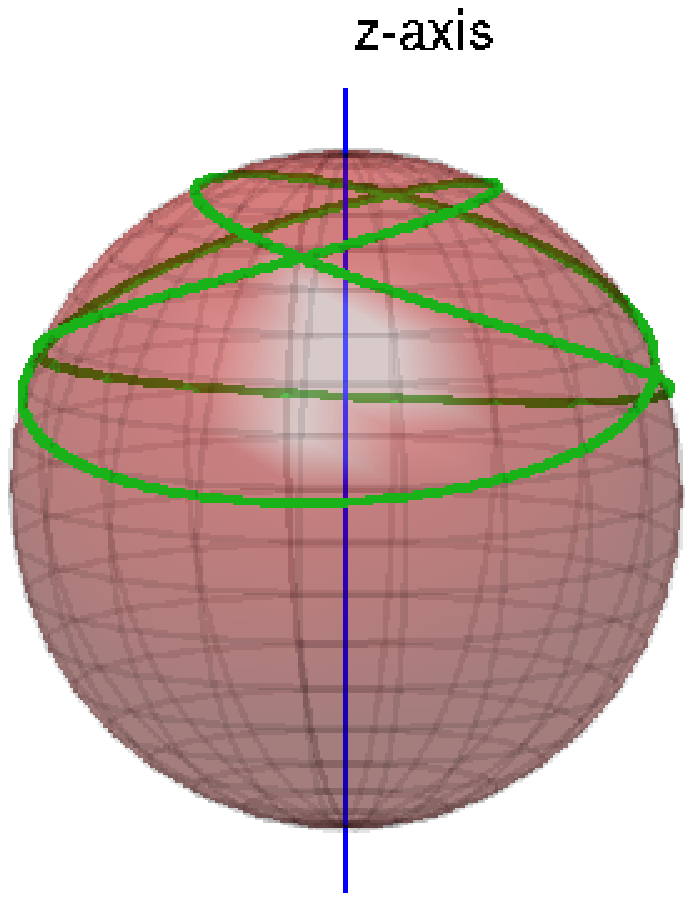} }
\put(40,29){{\large (d)}}
}
\end{picture} 
%============== 
\caption{Examples of trajectories of a single classical spin under the combined action of the static and the rf magnetic fields: (a) general case, open ergodic trajectory; (b) $\om=-\Ome$, closed trajectory; (c) $\om=-2\Ome$,
closed trajectory; (d) $\om=\frac{1}{2}\Ome$, closed trajectory.}
\label{fig:Trajectories}
\end{figure}
%%%%%%%%%%%%%%%%%%%%%%%%%%%%%%%%%%%%%%%%%%%%%%%%%%%%%%%%%%%%%%%%%%%

Let us now think about the averaging procedure in terms of classical spin trajectories. The typical situation corresponds to an irrational value of the ratio $\om/\Ome$. In this case, the trajectory of a classical spin that follows Eq.(\ref{eq:Spin-eom}) is open and ergodic on the spin sphere, as illustrated in Fig.~\ref{fig:Trajectories}(a). Averaging over such trajectories leads to the standard two-step truncation result. If $\om/\Ome$ is a rational number, then the individual spin trajectories become closed, and one can suspect that the averaging result is different from the irrational case. In reality, however, almost all rational values of $\om/\Ome$ lead to the same averaging results as the irrational values. In the case of like spins, the only exceptions are $\om/\Ome = \{ \pm 1/2, -1, -2 \}$ ---see Figs.~\ref{fig:Trajectories}(b-d).  The fact that no other rational numbers appear in this set is related to the fact that the interaction Hamiltonian $\Hdip$ is of the second order in terms of spin variables. If higher order spin couplings were involved, e.g. of the form $I_{ix} I_{jx} I_{ky}$, then other ratios with absolute values such as 1/3 or 3 may also appear in the above set.

\section{Results}
\label{results}

In this section, we present the time-averaged Hamiltonians separately for the couplings of ``like'' spins  (subsection~\ref{like}) and ``unlike'' spins  (subsection~\ref{unlike}). Like spins have the same gyromagnetic ratios and hence the same Larmor frequencies. Unlike spins have different gyromagnetic ratios and different Larmor frequencies. The Hamiltonians referring to the above two cases are marked by superscripts $^l$ and $^u$, respectively. The details of the calculations can be found in Ref.~\cite{supplement}.

\

\subsection{Like spins}
\label{like}

Away from the recoupling resonances, our averaging procedure reproduces the standard truncated Hamiltonian for like spins\cite{Slichter-90}:  
\begin{align}
   &\Hdipo=\sum_{i<j}^N J_{ij} \; \frac{1}{2} \; \bigl(3\cos^2 \alpha - 1\bigr)\; \left(1- \frac{3\rii{,z}}{\rii{}}\right)\nonumber\\ &\hspace{2cm}\times \Bigl[\Idr{iz}\Idr{jz}-\frac{1}{2}(\IIxyz{x}{x}{+}{y}{y})\Bigr] ,
\label{eq:Hdip0}
\end{align}
where $\alpha$ is given by Eq.(\ref{alpha}).

As already mentioned in Section~\ref{truncation}, the averaged Hamiltonian contains additional coupling terms in the case of recoupling resonances associated with ratios $\om/\Ome = \{ \pm 1/2, -1, -2 \}$.  According to Eq.(\ref{eq:Ome}), these ratios translate into the following expressions for $\om$ in terms of $\omo$ and $\omrf$:
\begin{align}
\om&= -\Ome : & \om&= - \frac{\omrf^2+\omo^2}{2\omo} ; \label{omOme} \\ 
\om&= - 2\Ome : &  \om&= \frac{- 4 \omo \pm 2 \sqrt{\omo^2-3\omrf^2}}{3} ; \label{om2Ome} \\ 
\om&=\pm \frac{1}{2}\Ome : & \om&=\frac{\omo \pm \sqrt{4 \omo^2 + 3 \omrf^2}}{3}. \label{2ommOme}
\end{align}
The upper (lower) sign before the square root in Eq.(\ref{2ommOme}) corresponds to the upper (lower) sign of the resonant condition $\om=\pm \frac{1}{2}\Ome$. In the case of Eq.(\ref{om2Ome}) both signs before the square root are realizable for the same condition $\om = - 2\Ome$ as long as $|\omo| \geq \sqrt{3} |\omrf| $.  When the value of $\omega$ following from Eq.(\ref{omOme}),(\ref{om2Ome}) or (\ref{2ommOme}) is negative, this indicates that frequency vector $\VEC{\omega}$ is anti-parallel to $\VEC{H}_0$. The resonance conditions $\om= \Ome$ and $\om= 2 \Ome$ are absent in the above list, because, after the substitution of $\Ome$ given by  Eq.(\ref{eq:Ome}) they give no real-valued solutions for $\omega$. 

Now we list the corresponding averaged Hamiltonians:
\begin{widetext}
\small
\begin{align}
& \om=-\Ome : &\HdipOme&= \Hdipo 
+\sum_{i<j}^N J_{ij} \; \left\{ \frac{3}{8}(\ca-1)^2 \; \Bigl[\frac{\rii[2]{,y}-\rii[2]{,x}}{\rii{}}(\IIxyz{x}{x}{-}{y}{y})+ \frac{ 2 \ \rii[]{,x}\rii[]{,y}}{\rii{}}(\IIxyz{x}{y}{+}{y}{x})\Bigr] \right. \nonumber \\
&&& + \left. \frac{3}{2}(\cos 2\alpha-\cos \alpha) \; 
\Bigl[-\frac{r_{ij,x}r_{ij,z}}{\rii{}}(\IIxyz{x}{z}{+}{z}{x})+ \frac{r_{ij,y}r_{ij,z}}{\rii{}}(\IIxyz{y}{z}{+}{z}{y})\Bigr] \right\} ;
\label{eq:Hdip(om=Ome)}\\[14pt]
&\om=-2\Ome : &\HdipTOme&=\Hdipo+\sum_{i<j}^N J_{ij} \; \frac{3}{4} \;[\sin 2 \alpha - 2 \sin\alpha] \; \Bigl[\frac{r_{ij,x}r_{ij,z}}{\rii{}}(\IIxyz{x}{x}{-}{y}{y})- \frac{ r_{ij,y} r_{ij,z}}{\rii{}}(\IIxyz{x}{y}{+}{y}{x})\Bigr] ;
\label{eq:Hdip(om=2Ome)}\\[14pt]
&\om=\pm\frac{1}{2}\Ome : &\HdipHOme&=\Hdipo + \sum_{i<j}^N J_{ij} \; \frac{3}{8} \; [\sin 2 \alpha \pm 2 \sin\alpha ] \; \Bigl[\frac{\rii[2]{,y}-\rii[2]{,x}}{\rii{}}(\IIxyz{x}{z}{+}{z}{x})\mp \frac{2 \ \rii[]{,x}\rii[]{,y}}{\rii{}}(\IIxyz{y}{z}{+}{z}{y})\Bigr]. \label{eq:Hdip(om=1/2Ome)}
\end{align}
\end{widetext}
The upper (lower) sign of the resonance condition of Eq.(\ref{eq:Hdip(om=1/2Ome)}) corresponds to the upper (lower) sign in the expression for $\HdipHOme$.

We note that the standard truncated Hamiltonian $\Hdipo$ conserves the $z'$-projection of the total spin polarization $\sum_i I_{iz}'$ in the double-rotating reference frame. This total spin component still relaxes on the time scale of the order of $T_1$ because of the interaction with electronic spins or phonons. We call this ``normal longitudinal relaxation.''  In contrast, the recoupled terms appearing in Eqs.(\ref{eq:Hdip(om=Ome)}-\ref{eq:Hdip(om=1/2Ome)}) do not conserve $\sum_i I_{iz}'$, which means that all three projections of the total spin polarization decay to zero in both the double-rotating and the laboratory reference frames on a time scale, which is, in general, much faster than $T_1$. We call this ``anomalous longitudinal relaxation.''

The recoupled terms in Eqs.(\ref{eq:Hdip(om=Ome)}-\ref{eq:Hdip(om=1/2Ome)}) are comparable with $\Hdipo$, when  $H_1\sim H_0$. In such a case, the time scale of the anomalous longitudinal relaxation is expected to be of the order of $T_2$, which is clearly much faster than $T_1$. Therefore, the recoupling resonances should be readily identifiable in an  experiment of the type proposed in Section~\ref{experiment}.

\begin{figure} \setlength{\unitlength}{0.1cm} 
%=======================================================================

\begin{picture}(80, 57)
{ 
\put(-5,0){ \epsfxsize= 1.0\linewidth \epsfbox{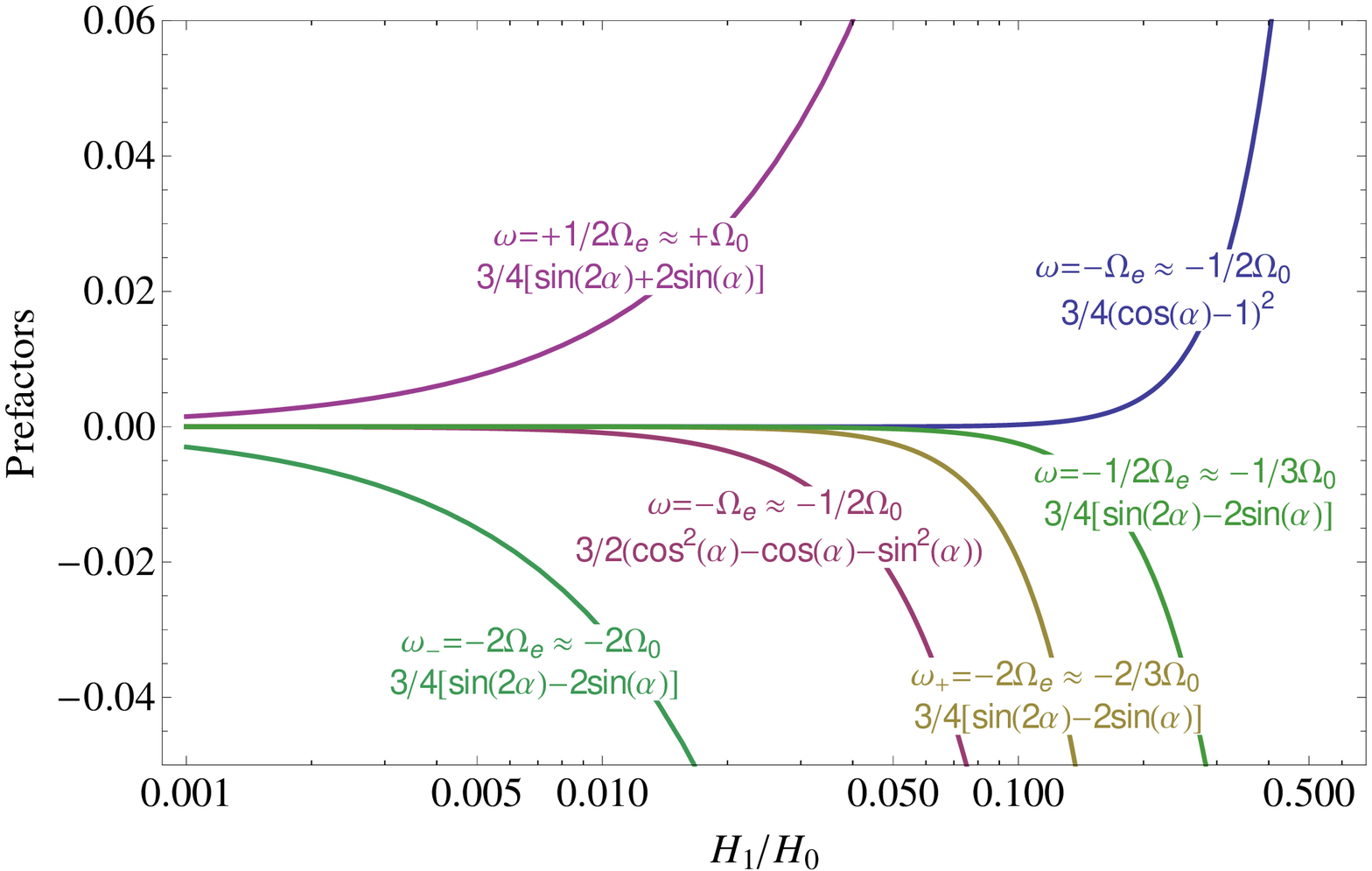} }
}
\end{picture}
%============== 
\caption{(Color online) Plots of the $\alpha$-dependent prefactors of the recoupled terms in the like-spin Hamiltonians (\ref{eq:Hdip(om=Ome)}-\ref{eq:Hdip(om=1/2Ome)}) as functions of the ratio $H_1/H_0$ for $H_1< 0.5 H_0$. The legend for each line indicates the corresponding resonant condition and the $\alpha$-dependence of the prefactor being plotted. Each resonant condition is given both in its exact form (\ref{omOme}-\ref{2ommOme}) and in the approximate form for the case $H_1 \ll H_0$. The conditions $\omega_{+}=-2\Omega_e$ and $\omega_{-}=-2\Omega_e$ correspond to the recoupling resonances (\ref{om2Ome}) with signs ($+$) or ($-$), respectively.}
\label{fig:Pref}
\end{figure}

In the case of $H_1\ll H_0$, the recoupled terms can still induce the anomalously fast longitudinal relaxation, but this case requires a more detailed discussion. We first note that all recoupled terms in Eqs.(\ref{eq:Hdip(om=Ome)}-\ref{eq:Hdip(om=1/2Ome)}) are equal to zero, when $\alpha = 0$, which, according to Eq.(\ref{alpha}), corresponds to $H_1=0$. When $H_1\ll H_0$, the recoupled terms are much smaller than the terms in $\Hdipo$, in agreement with the standard truncation result of Redfield\cite{Redfield-55,Abragam-61,Slichter-90}. The values of the $\alpha$-dependent prefactors of the recoupled terms in the Hamiltonians (\ref{eq:Hdip(om=Ome)}-\ref{eq:Hdip(om=1/2Ome)}) are plotted as functions of $H_1/H_0$ in Fig.\ref{fig:Pref} for $H_1/H_0 \leq 0.5 $.
The strongest of these terms are of the order of $H_1/H_0$. They correspond to the resonances $\omega \approx \Omega_0$  and $\omega \approx - 2 \Omega_0$ obtained, respectively, from Eq.(\ref{om2Ome}) with plus sign and Eq.(\ref{2ommOme}) with minus sign in the limit $H_1/H_0 \rightarrow 0$. (In our sign convention, the usual single-spin NMR resonance is located at $\omega \approx -\Omega_0$.)

In general, the recoupled terms linear in $H_1/H_0$ should induce a longitudinal relaxation on the time scale of the order of $T_2 (H_0/H_1)^2$, which can still be much shorter than $T_1$ and thus represent a clear indication of a recoupling resonance. The above estimate is of the second order in terms of $H_0/H_1$, because the first-order effect of the recoupled terms should be averaged out by the spin motions due to the much stronger secular terms. 
In view of this consideration, the observability of the recoupling resonances can be significantly improved by suppressing the secular Hamiltonian $\Hdipo$. If this is done,  the timescale of the longitudinal relaxation induced by the recoupled terms can be estimated as $T_2 (H_0/H_1)$, i.e. it becomes much faster. 

The suppression of $\Hdipo$ makes the observable effects of the recoupling resonances more dramatic not only in the limit of $H_1 \ll H_0$, but also in the general case of $H_1 \sim H_0$. It can be realized, for example, as follows. 

A given recoupling resonance imposes a constraint on the three parameters $H_0$, $H_1$ and $\omega$, which leaves a freedom to impose one additional relationship. This freedom can be used to impose the magic angle condition \mbox{$1-3 \cos^2 \alpha = 0$}.  However, this condition together with any of the like-spin recoupling resonances fixes the ratio $H_1/H_0$ to be of the order of 1, which means that it cannot be used if the experiment is limited to the regime $H_1 \ll H_0$. 

Another resource for suppressing $\Hdipo$ is to exploit the spatially dependent factor $1-\frac{3\rii{,z}}{\rii{}}$. For example, if the recoupling resonances are to be investigated in  CaF$_2$, where $^{19}$F nuclei form a simple cubic lattice, it is better to use the static field along the [111] direction of that lattice. In such a case case, $1-\frac{3\rii{,z}}{\rii{}} =0$ for all six nearest neighbors of each spin, i.e. the secular interaction with these neighbors is suppressed, while the recoupled terms remain large.

Of course, the most effective way to suppress the secular interactions for all pairs of interacting spins is to use the MAS technique\cite{Slichter-90}. This technique completely averages out all factors $1-\frac{3\rii{,z}}{\rii{}}$, when $\cos^2 \Theta = 1/3$, where $\Theta$ is the ``magic'' angle between the spinning axis and the static magnetic field. Spinning at the magic angle, however, does not average out completely the factors $r^2_{ij,y}-r^2_{ij,x}$, $r_{ij,x}r_{ij,y}$, or  $r_{ij,x}r_{ij,z}$ that enter the recoupled Hamiltonians.

%%%%%%%%%%%%%%%%%%%%%%%%%%%%%%%%%%%%%%%%%%%%%%%%%%%%%%%%%%%%%%%%%%%
%%%%%%%%%%%%%%%%%%%%%%%%%%%%%%%%%%%%%%%%%%%%%%%%%%%%%%%%%%%%%%%%%%%

\subsection{Unlike spins}
\label{unlike}

In this subsection, we consider coupling between two unlike spin species. We will use lattice index $i$ to label one of these species and index $j$ to label the other. We reintroduce the labeling of gyromagnetic ratios $\gamma_i$ and $\gamma_j$, and the corresponding Larmor frequencies $\Omega_{0i} = \gamma_i H_0$ and $\Omega_{0j} = \gamma_j H_0$. The two spin species are now to be described in two different double-rotated reference frames. The effective Larmor frequencies $\Omega_{ei}$, $\Omega_{ej}$ and the corresponding tilting angles $\alpha_i$, $\alpha_j$ are then defined by adding the indices $i$ or $j$ to variable $\Omega_e$, $\Omega_0$, $\omega_1$ and $\alpha$ in Eqs.(\ref{eq:Ome}) and (\ref{alpha}). 

As in the case of like spins, our averaging procedure for unlike spins recovers the standard truncated Hamiltonian\cite{Slichter-90} away from the recoupling resonances: 
\begin{align}
  \HDu{0}&=\sum_{\langle i,j \rangle}\Rfactor\frac{1}{2}[3\cos\alpha_i \cos \alpha_j -\cos(\alpha_i - \alpha_j)]\nonumber\\
  &\hspace{1cm}\times \left(1- \frac{3\rii{,z}}{\rii{}}\right) \Idr{iz}\Idr{jz},\label{eq:H2dip0}
\end{align}
where the sum $\sum_{\langle i,j\rangle} $ includes all pairs of unlike spins. 

It should be mentioned here, that, in the very special case of $\gamma_i = -\gamma_j$ and $\omega = 0$, the secular Hamiltonian $\HDu{0}$ should also include the so-called ``double-flip'' terms $I'_{i+} I'_{j+}$ and $I'_{i-} I'_{j-}$, where $I'_{i\pm} = I'_{ix} \pm I'_{iy}$, etc. These terms are preserved by the averaging, because their counterparts $I_{i+} I_{j+}$ and $I_{i-} I_{j-}$  in the original full Hamiltonian (\ref{Htot}) now conserve the Zeeman energy.  This case can be mapped onto a somewhat unusual problem of like spins by changing the sign of all y- and z- spin projections for one of the two spin species, e.g. $I_{jy} \rightarrow -I_{jy}$ and $I_{jz} \rightarrow -I_{jz}$. However, when $\omega \neq 0$, the two spin species in the transformed problem will experience rf-fields rotating in the opposite directions, which means that the transformed problem is not equivalent to the problem of like spins in a single rotating field. 

Another special case is $\gamma_j \rightarrow  0$. It implies that the terms of the type $I_{iz} I_{j\pm}$ in the full Hamiltonian (\ref{Htot}) become secular, and, as a result, terms $I'_{iz} I'_{j\pm}$ may appear in $\HDu{0}$.  Such a limit is not realizable physically in the case of magnetic dipole interaction, because $\gamma_j = 0$ simultaneously suppresses the interaction itself.  

The above special cases should be kept in mind in the analysis of the parameter dependence of the recoupling resonances.

There exist two kinds of ``unlike'' recoupling resonances, namely, the resonances involving the motion of only one spin species, and the resonances involving both spin species. The resonances of the first kind are the following:
\begin{align}
 \om&=-\Ome[i] \; : & \om&= - \frac{\omrf[i]^2+\omo[i]^2}{2\omo[i]}, \label{eq:om=Omei} \\ 
 \om&=\pm\frac{1}{2}\Ome[i] \; : & \om&=\frac{-\omo[i]\pm\sqrt{4\omo[i]^2+3\omrf[i]^2}}{3}, \label{eq:om=1/2Omei} 
\end{align}
and also those obtained from Eq.(\ref{eq:om=Omei},\ref{eq:om=1/2Omei}) by replacing index $i$ with index $j$. The resonances of the second kind are
\begin{align}
 \om&=(-1)^n\frac{1}{2}(\Ome[i]\pm\Ome[j]) \label{eq:om=1/2(Ome1+Ome2)}, \\
 \om&=(-1)^n(\Ome[i]\pm\Ome[j]), \label{eq:om=(Ome1+Ome2)}
\end{align}
where variable $n$ takes values 0 or 1 and thus controls the sign in front of the entire expression. Resonant conditions (\ref{eq:om=1/2(Ome1+Ome2)},\ref{eq:om=(Ome1+Ome2)}) lead to fourth-order polynomial equations with respect to $\omega$, which, in principle can be solved analytically. However, in view of the cumbersome character of the resulting formulas, we chose to investigate the solutions numerically. Several plots illustrating the character of these solutions are presented in Fig.~\ref{fig-pref-main} and further in Fig.~\ref{fig-pref-app} of Appendix~\ref{app}.  As evident from these plots, it cannot be guaranteed that each of the eight conditions envisioned by Eqs.(\ref{eq:om=1/2(Ome1+Ome2)},\ref{eq:om=(Ome1+Ome2)}) is realizable for a given set of values of $\gamma_i$, $\gamma_j$, $H_0$ and $H_1$.  For example, if $\gamma_i,\gamma_j > 0$, then no value of $\omega$ exists that would satisfy the condition
$\omega = 1/2 (\Ome[i]+ \Ome[j])$. 

The time-averaged Hamiltonians corresponding to the recoupling resonances (\ref{eq:om=1/2(Ome1+Ome2)},\ref{eq:om=(Ome1+Ome2)}) are the following:
\begin{widetext}
\small
\begin{align}
&\om=-\Ome[i]\;:&\HDu{1}&=\HDu{0}+\sum_{\langle i,j\rangle} J_{ij} \frac{3}{2} \left[\cos(\alpha_i+\alpha_j)-\cos\alpha_j \right] \; \Bigl[-\frac{r_{ij,x}r_{ij,z}}{r_{ij}^2}\I_{ix}\I_{jz}+ \frac{r_{ij,y}r_{ij,z}}{r_{ij}^2}\I_{iy}\I_{jz}\Bigr] , \label{eq:HDu1}\\[14pt]
&\om=\pm\frac{1}{2}\Ome[i]\;:&\HDu{2}&=\HDu{0}+\sum_{\langle i,j\rangle} J_{ij}\; \frac{3}{4} \; [\sin\alpha_j(\cos\alpha_i\pm 1)] \; \Bigl[\frac{r_{ij,y}^2-r_{ij,x}^2}{r_{ij}^2}\I_{ix}\I_{jz}\mp \frac{2 \ r_{ij,x}r_{ij,y}}{r_{ij}^2}\I_{iy}\I_{jz}\Bigr] , \label{eq:H2dip(om=Ome)}\\[14pt]
&\om=(-1)^n(\Ome[i]\pm\Ome[j])\;:&\HDu{3}&=\HDu{0}+\sum_{\langle i,j\rangle} J_{ij} \frac{3}{4} \; \left[\sin(\alpha_i+\alpha_j)\pm (-1)^n\sin\alpha_i+ (-1)^n\sin\alpha_j\right]\nonumber\\
&&&\hspace{2cm}\cdot\left[\frac{r_{ij,x}r_{ij,z}}{r_{ij}^2} \left(\IIxyz{x}{x}{\mp}{y}{y}\right)+(-1)^n\frac{r_{ij,y}r_{ij,z}}{r_{ij}^2} \left(\pm\IIxyz{x}{y}{+}{y}{x}\right)\right] ,\label{eq:H2dip(om=(Ome1+Ome2))}\\[14pt]
&\om=(-1)^n\frac{1}{2}(\Ome[i]\pm\Ome[j])\;:&\HDu{4}&=\HDu{0} +\sum_{\langle i,j\rangle} J_{ij} \frac{3}{8} \; \left[\cos\alpha_i+(-1)^n\right]\left[\cos\alpha_j\pm (-1)^n\right]\nonumber\\ 
&&&\hspace{2cm}\cdot\left[\frac{r_{ij,y}^2-r_{ij,x}^2}{r_{ij}^2} \left(\IIxyz{x}{x}{\mp}{y}{y}\right)-(-1)^n\frac{2\;r_{ij,x}r_{ij,y}}{r_{ij}^2} \left(\pm\IIxyz{x}{y}{+}{y}{x}\right)\right] .\label{eq:H2dip(om=1/2(Ome1+Ome2))}
\end{align}
\end{widetext}
The upper (lower) sign of the recoupling resonances of Eqs.(\ref{eq:H2dip(om=(Ome1+Ome2))},\ref{eq:H2dip(om=1/2(Ome1+Ome2))}) corresponds to the upper (lower) signs in the corresponding Hamiltonians. In the case of a small mismatch of the above resonant conditions, the additional term (\ref{DeltaH}) for one or both spin species should be included. In the case of resonances (\ref{eq:H2dip(om=(Ome1+Ome2))},\ref{eq:H2dip(om=1/2(Ome1+Ome2))}), one, in fact, has the freedom of distributing the mismatch between the two spin species.

%%%%%%%%%%%%%%%%%%%%%%%%%%%%%%%%%%%%%%%%%%%%%%%%%%%%%%%%%%%%%%%%%%%
\begin{figure} \setlength{\unitlength}{0.1cm} 
%=======================================================================

\begin{picture}(80, 138)
{ 
\put(21,0){$\gamma_j/\gamma_i$}
\put(61,0){$\gamma_j/\gamma_i$}
\put(0,0){ \epsfxsize= 4cm \epsfbox{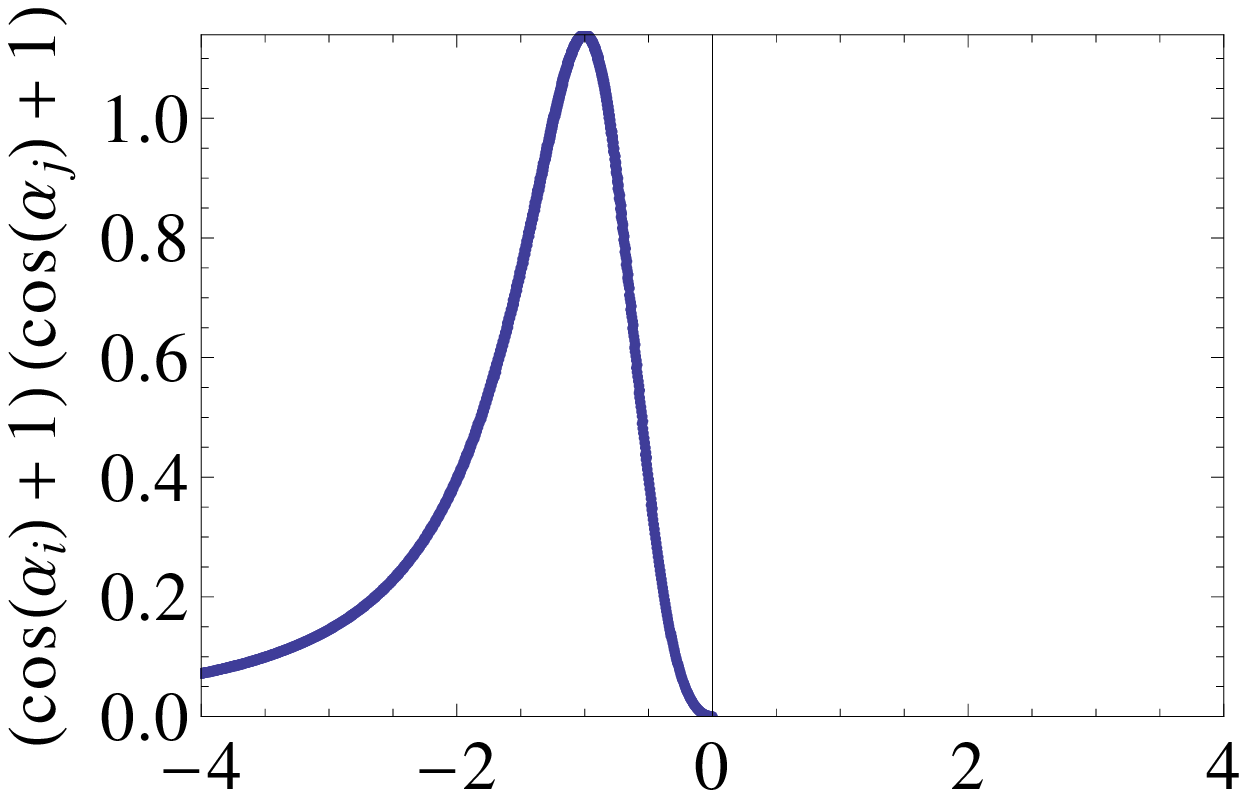} }
\put(41,0){ \epsfxsize= 4cm \epsfbox{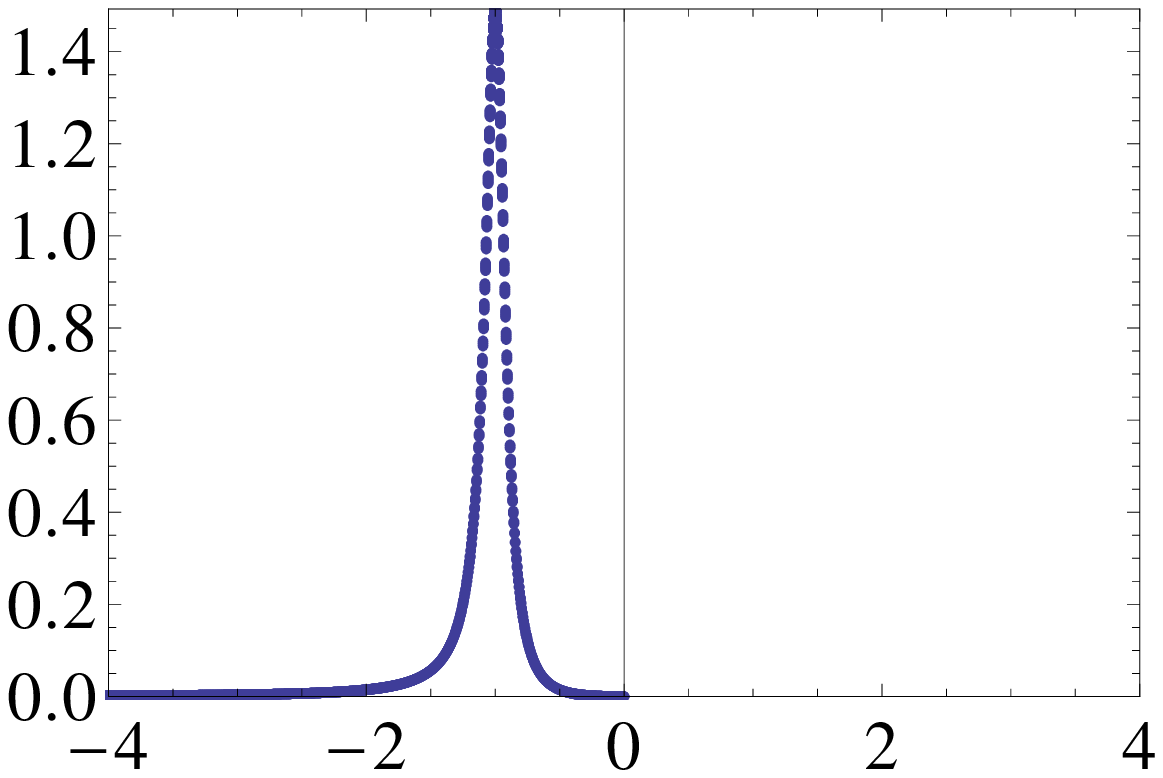} }
\put(0,27){ \epsfxsize= 4cm \epsfbox{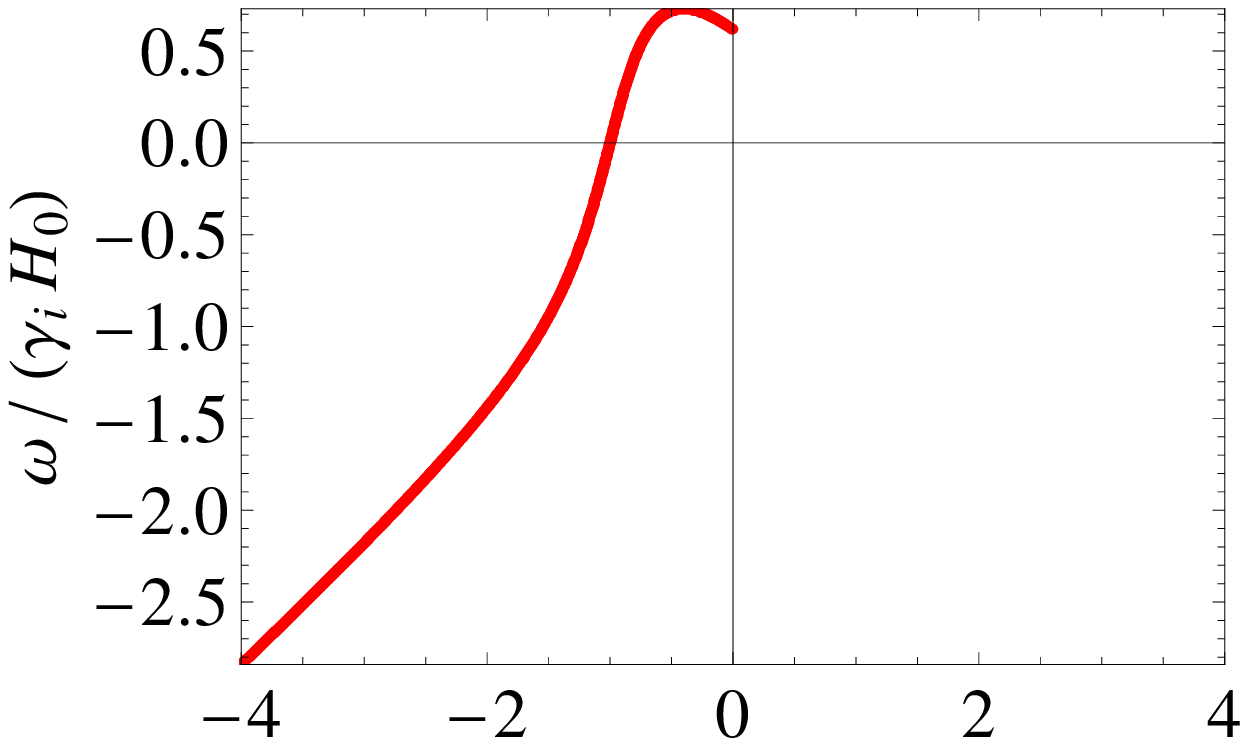} }
\put(40,27){ \epsfxsize= 4cm \epsfbox{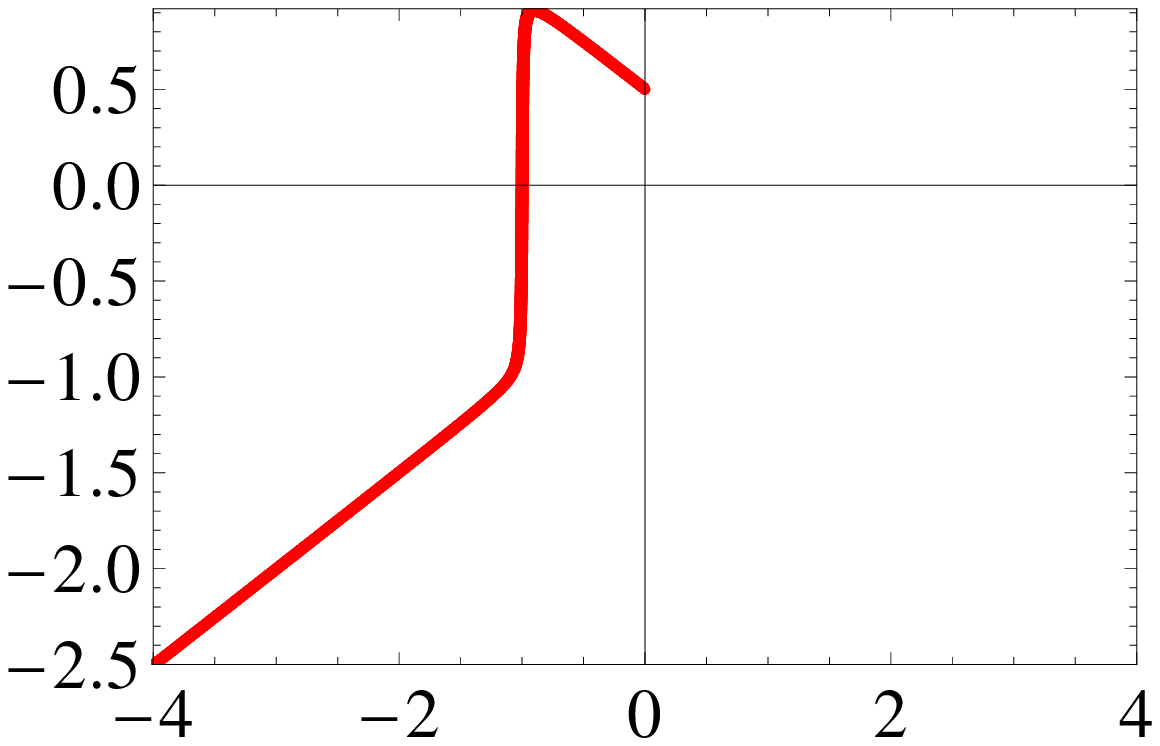} }
\put(30,58){$\omega=\frac{1}{2}(\Omega_{ei}+\Omega_{ej})$}
\put(-1,64){ \epsfxsize= 4cm \epsfbox{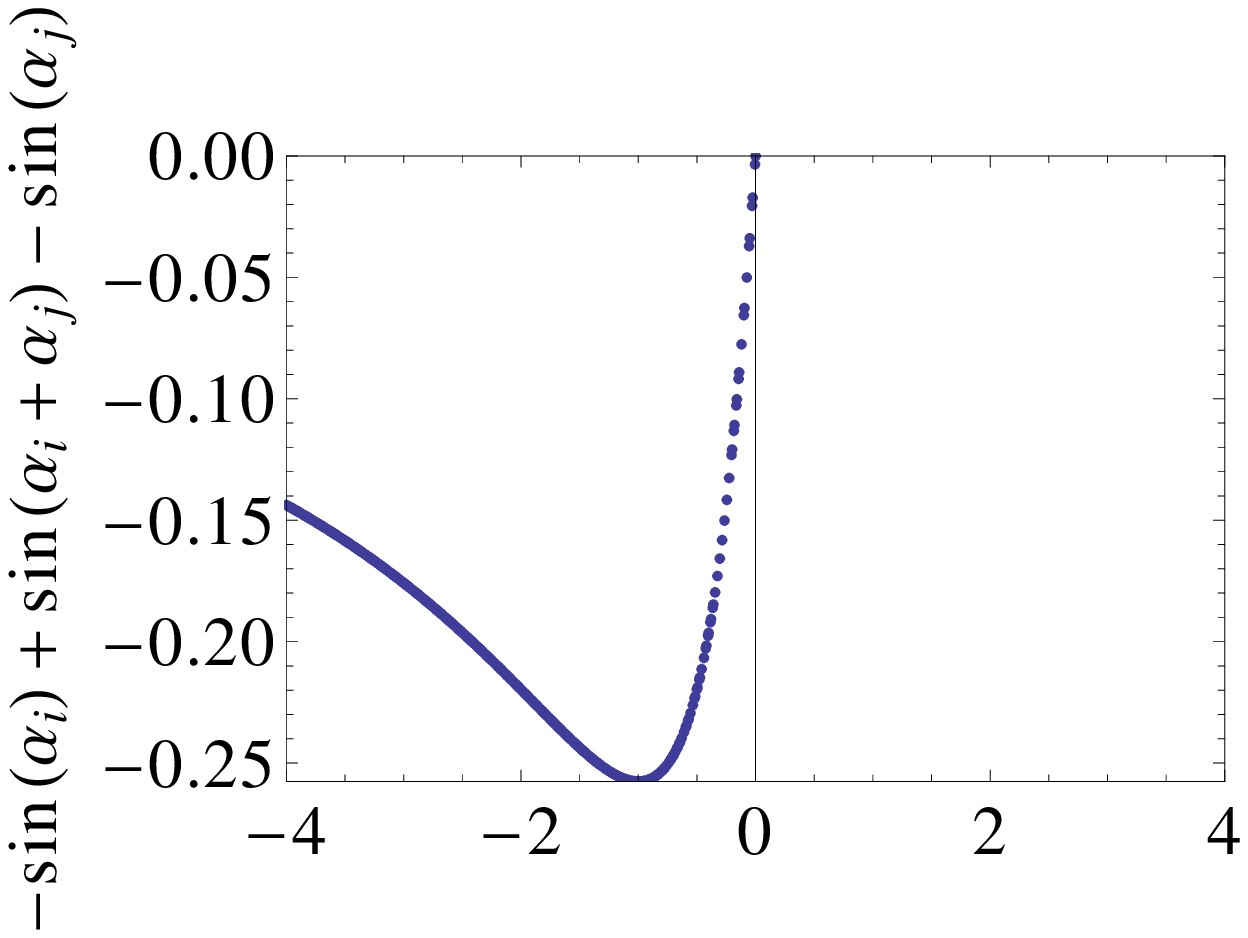} }
\put(39.5,64){ \epsfxsize= 4cm \epsfbox{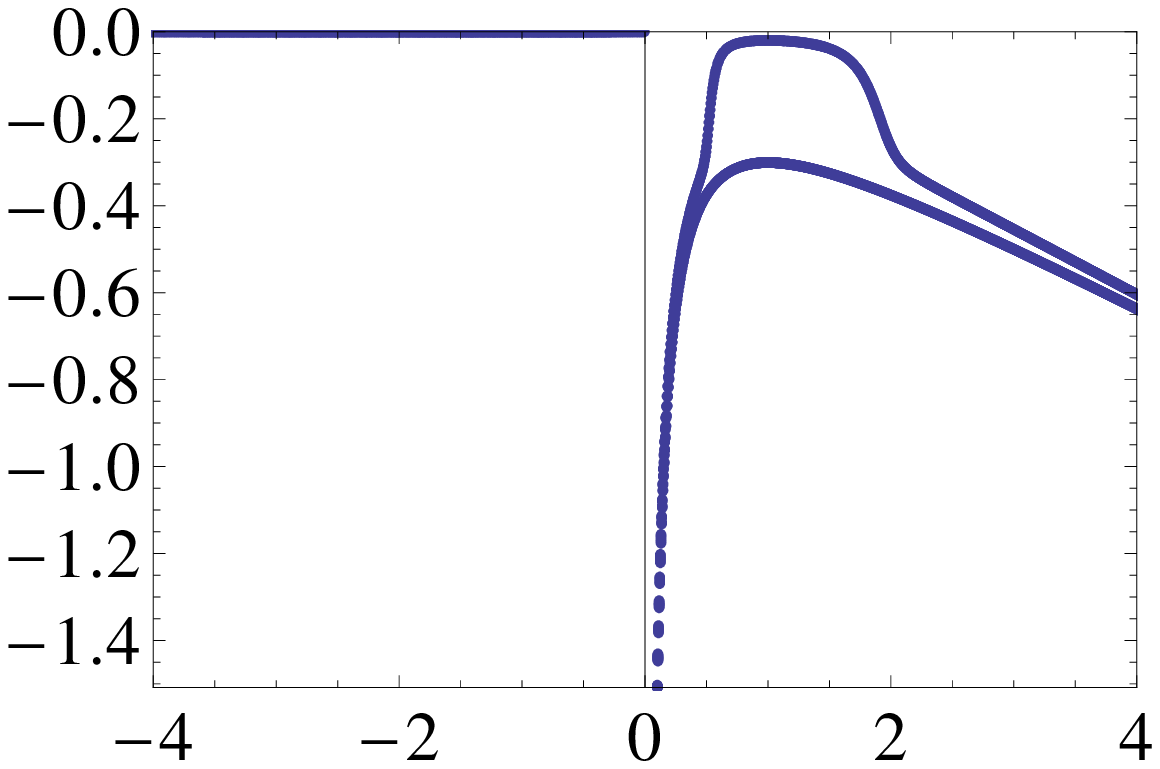} }
\put(0,90){ \epsfxsize= 4cm \epsfbox{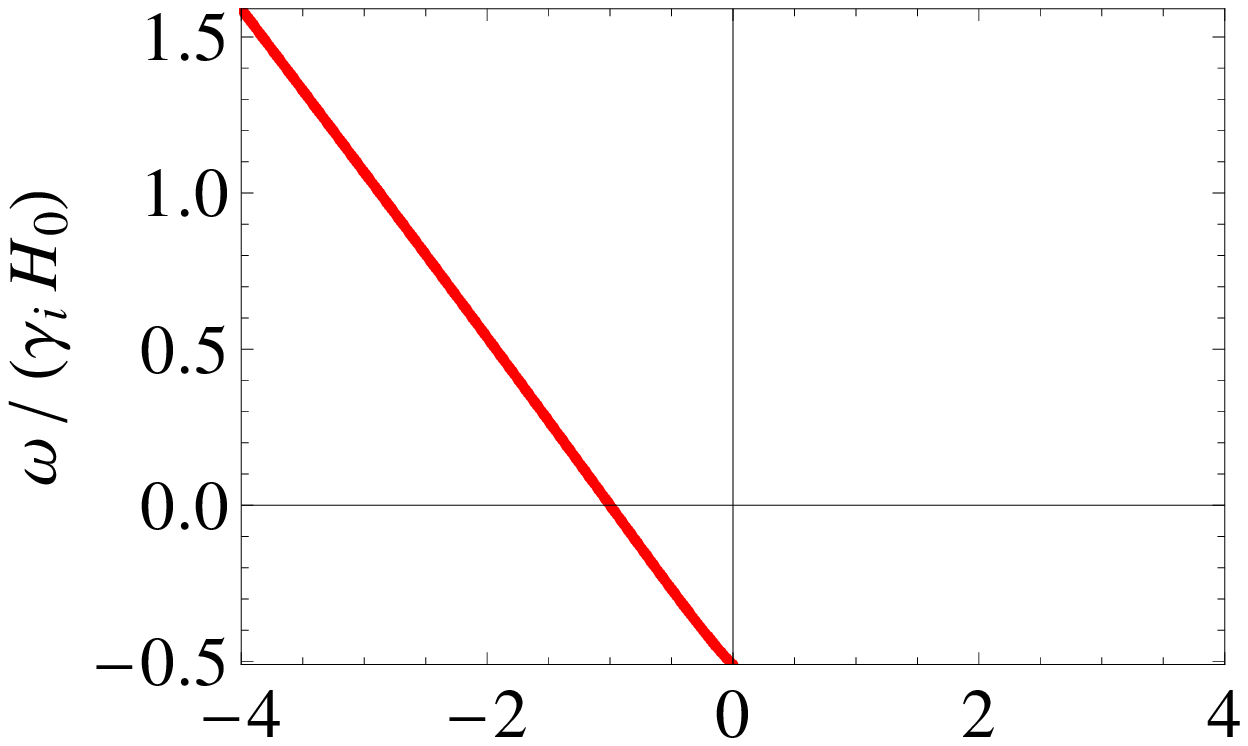} }
\put(40,90){ \epsfxsize= 4cm \epsfbox{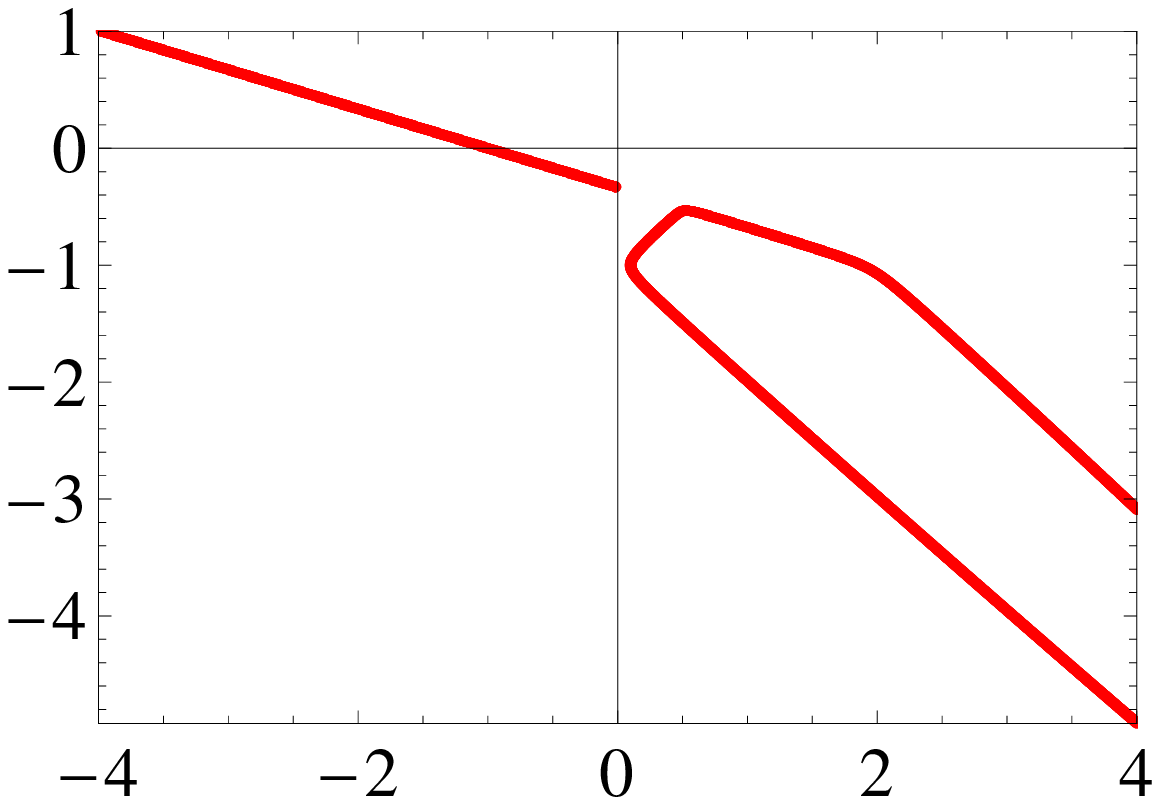} }
\put(30,121){$\omega=-(\Omega_{ei}+\Omega_{ej})$}

\put(15,128){$\frac{H_1}{H_0}=0.9$}
\put(55,128){$\frac{H_1}{H_0}=0.1$}

\put(0,122){\bf (a)}
\put(0,59){\bf (b)}
}
\end{picture}
%============== 
\caption{(Color online) Examples of recoupling resonances for unlike spins. Plots of the resonant values of $\omega$ and the prefactors in the recoupled Hamiltonians (\ref{eq:H2dip(om=(Ome1+Ome2))},\ref{eq:H2dip(om=1/2(Ome1+Ome2))}) as functions of $\gamma_j/\gamma_i$  for two resonant conditions indicated above the respective plots. All examples presented are obtained numerically for either $H_1/H_0=0.9$ or $H_1/H_0=0.1$. Each column of plots corresponds to the ratio $H_1/H_0$  indicated above it. The plot lines consist of the actual computed points and thus in some cases have the appearance of dotted lines. 
The results for $H_1/H_0=0.1$ are qualitatively representative of the limit $H_1 \ll H_0$.
The analogous plots for the six remaining resonant conditions associated with Eqs.(\ref{eq:H2dip(om=(Ome1+Ome2))},\ref{eq:H2dip(om=1/2(Ome1+Ome2))}) are given in Appendix~\ref{app}.}
\label{fig-pref-main}
\end{figure}
%%%%%%%%%%%%%%%%%%%%%%%%%%%%%%%%%%%%%%%%%%%%%%%%%%%%%%%%%%%%%%%%%%%

The standard truncated Hamiltonian (\ref{eq:H2dip0}) conserves both $\sum_i I'_{iz}$ and $\sum_j I'_{jz}$, while each of the recoupled Hamiltonians \eqref{eq:HDu1}-\eqref{eq:H2dip(om=1/2(Ome1+Ome2))} violates the above conservation for at least one of the two spin species. We observe, however, that the total $z'$-polarization of the both spin species, i.e. $\sum_i I'_{iz} + \sum_j I'_{jz}$, is still conserved by the recoupled Hamiltonians given by Eq.(\ref{eq:H2dip(om=(Ome1+Ome2))}) for $\om=\pm(\Ome[i]-\Ome[j])$ and by Eq.(\ref{eq:H2dip(om=1/2(Ome1+Ome2))}) for $\om=\pm \frac{1}{2}(\Ome[i]-\Ome[j])$, because the terms in these Hamiltonians have the character of the so-called ``flip-flops."

In general, the recoupled Hamiltonians (\ref{eq:om=Omei}-\ref{eq:om=(Ome1+Ome2)}) are comparable to the standard truncated Hamiltonian (\ref{eq:H2dip0}), when $H_1$ is comparable to $H_0$ --- see plots in Fig.~\ref{fig-pref-main} for $H_1/H_0 = 0.9$. If $H_1 \ll H_0$, then the recoupled terms are normally small, because Eq.(\ref{alpha}) gives the values of $\alpha_i$ and $\alpha_j$ either close to 0 or to $\pi$ --- Figs.~\ref{fig-pref-app}(c,d,f) for $H_1/H_0 = 0.1$.

However, as the plots in Figs.~\ref{fig-pref-main} and \ref{fig-pref-app}(a,b,e) for $H_1/H_0 = 0.1$ indicate, and our approximate calculations confirm, there are interesting and potentially useful exceptions, when the angle-dependent prefactors of the recoupled Hamiltonians remain comparable to 1 even though $H_1 \ll H_0$. The first exception corresponds to the condition  $\gamma_j/\gamma_i \approx -1$ for the recoupling resonances $\om=\pm 1/2 (\Ome[i]-\Ome[j])$ and $\om= 1/2 (\Ome[i]+\Ome[j])$. In these cases, the recoupled terms in Hamiltonians (\ref{eq:H2dip(om=1/2(Ome1+Ome2))}) remain large as long as $|\gamma_j/\gamma_i + 1| \lesssim H_1/H_0$. [In the case of resonance $\om= 1/2 (\Ome[i]+\Ome[j])$, one should also be mindful of adding the secular double-flip contribution mentioned after Eq.(\ref{eq:H2dip0}).]   This group of resonances can thus be exploited for pairs of nuclei satisfying the condition \mbox{$\gamma_i\approx -\gamma_j$}, such as e.g. $^{129}$Xe and $^{23}$Na. The second exception corresponds to  the condition of $\gamma_j/\gamma_i \approx 0$  (or  $\gamma_i/\gamma_j \approx 0$) for the resonances $\om=\pm(\Ome[i]+\Ome[j])$. The corresponding recoupled Hamiltonians (\ref{eq:H2dip(om=(Ome1+Ome2))}) are large, as long as $|\gamma_j/\gamma_i| \lesssim H_1/H_0$. These recoupling resonances can be very useful for the purposes of cross-polarization, when the gyromagnetic ratio of one of the two nuclei is much smaller than the other.   

Finally, we mention that, as in the case of like spins, the suppression of the secular interaction terms between unlike spins by matching the magic angle condition or by the MAS technique is an additional resource for observing the recoupling resonances. The only difference in the present case is that the magic angle condition involves two angles: $3\cos\alpha_i \cos \alpha_j -\cos(\alpha_i - \alpha_j) = 0$.

\section{Discussion}
\label{discussion}

\subsection{Experimental realizability}
\label{experiment}

The standard truncated Hamiltonians (\ref{eq:Hdip0},\ref{eq:H2dip0}) conserve the $z'$-projection of the total spin polarization for each of the available spin species in their respective double-rotated reference frames.  The most obvious qualitative effect of the recoupled Hamiltonians is that they do not conserve this total $z'$-projection for at least one of the spin species.  
This difference can be tested experimentally. In the case of like spins, it will lead to the anomalously fast longitudinal relaxation discussed in Section~\ref{like}. In the case of unlike spins, it may, in addition, lead to an anomalous cross-polarization effect.

To be specific, a possible way to observe the anomalous longitudinal relaxation can consist of the following steps: (i) polarizing the system in the static magnetic field; (ii) turning on the rf field (preferably, adiabatically) to match one of the recoupling resonance conditions; (iii) waiting enough time for the additional interaction terms to relax $\sum_i I_{iz}'$ (the waiting time should still be much shorter than $T_1$); and, finally, (iv) suddenly turning off the rf field. No free induction decay signal should be observable in this case. In contrast, away from the recoupling resonances, a free induction decay signal of finite intensity should always be observable for waiting times of the order of $T_1$.
 
The requirements for the above kind of experiments are the following: (1) The static and the rf fields should be strong enough to ensure $\omo, \omrf \gg 1/T_2$. (2) As discussed in the context of Eq.(\ref{DeltaH}), the recoupling resonances should be matched to an accuracy of roughly $\frac{1}{T_2}$. (3) The static and the rf magnetic fields should be relatively homogeneous; otherwise the recoupling conditions cannot be satisfied over the entire sample. (4) The strength of the recoupled terms must be such that the timescale associated with the anomalous relaxation of the longitudinal magnetization is faster than $T_1$ (see the end of Section~\ref{like}). (5) Finally, in the case of $H_1 \sim H_0$, a real circularly polarized rf field should be used, as done, e.g., in Ref.\cite{Sakellariou-05}.  

Now we discuss the observability of the recoupling resonances in the NMR setting that uses a linearly polarized rf field in the regime of $H_1 \ll H_0$. The latter condition extends to the experiments in larger static fields, which means stronger NMR signals.

The Hamiltonian averaging procedure used in this paper cannot be straightforwardly extended to linearly polarized rf fields, because  a closed analytical solution for the spin motion is not available in such a case. In particular, the usual assumption that one can neglect one of the two counter-rotating components contributing to the linearly polarized rf field is not justified when $H_1 \sim H_0$. However, when $H_1 \ll H_0$, the case of linearly polarized rf field should be treatable by combining our truncation procedure with the perturbation expansion\cite{Bloch-40,Abragam-61}. In this limit, the presence of two rotating components may lead to additional recoupling resonances, but the resonances obtained in the present paper for a single rotating component of the same frequency should also be present. We further expect that the same recoupling resonance  in the rotating and the linearly polarized settings is generically characterized by the same recoupled Hamiltonian, when the leading order of the expansion of this Hamiltonian in terms of $H_1/H_0$ is linear. 

At small values of $H_1/H_0$, the recoupling resonances satisfying the above linearity requirement are also the strongest and thus the most promising in terms of experimental observation. As discussed in Section~\ref{like} for the case of like spins, the linearity with respect to small $H_1/H_0$  is exhibited by some but not all recoupled Hamiltonians. We also note in this context, that an experiment with small linearly polarized fields should avoid the interplay with the secondary single-spin resonances described by Winter\cite{Winter-55,Abragam-61}.)

In addition to the possibility of using stronger static fields and linearly polarized rf fields, the regime of $H_1 \ll H_0$ has two other advantages. First, the rf field homogeneity requirement is less stringent, when $H_1 \ll H_0$, because what matters is the homogeneity of the effective frequencies $\Omega_{ei}$, to which the leading rf field contribution is only of the order of $(H_1/H_0)^2$. Second, in the case, when the secular part of the interaction is externally suppressed, the smallness of $H_1/H_0$ implies that the recoupling resonances can be resolved with higher accuracy (see the discussion in Section~\ref{truncation}).

\subsection{Possible applications}
\label{applications}

{\it Fundamental studies of spin-spin relaxation:}
Perhaps, the most direct use of the tunable Hamiltonians (\ref{eq:Hdip(om=Ome)}-\ref{eq:Hdip(om=1/2Ome)}, \ref{eq:HDu1}-\ref{eq:H2dip(om=1/2(Ome1+Ome2))}) is to conduct fundamental experimental studies of nuclear spin-spin relaxation in solids, which is still not completely understood, in a much broader range of parameters than those accessible with the standard truncated Hamiltonians.

{\it Cross-polarization:} 
In the case of unlike spins, the recoupled Hamiltonians (\ref{eq:H2dip(om=(Ome1+Ome2))}-\ref{eq:H2dip(om=1/2(Ome1+Ome2))}) for the resonances $\omega=\pm(\Ome[i]-\Ome[j])$ and $\omega=\pm\frac{1}{2}(\Ome[i]-\Ome[j])$ can be used to cross-polarize different spin species. As mentioned in Section~\ref{unlike}, these Hamiltonians contain ``flip-flop'' terms $\IIxyz{x}{y}{-}{y}{x}$ and $\IIxyz{x}{x}{+}{y}{y}$, which can transfer polarization from one spin species to the other. Such a transfer would require an rf field rotating with a single frequency, as opposed to the standard Hartmann-Hahn cross polarization routine \cite{Hartmann-62}, which involves two rf-field frequencies. The two spin species can also be cross-polarized, albeit less efficiently, using the transient effect of the ``double-flip'' terms $\IIxyz{x}{y}{+}{y}{x}$ and $\IIxyz{x}{x}{-}{y}{y}$ for resonances $\omega=\pm(\Ome[i]+\Ome[j])$ and $\omega=\pm\frac{1}{2}(\Ome[i]+\Ome[j])$.
The suppression of the secular terms by either the magic angle condition or the MAS technique, as discussed in section~\ref{unlike}, can enhance the efficiency of both the flip-flop-based and the double-flip-based cross-polarization processes. In the latter case this enhancement should be particularly significant.  

{\it Structure determination:}
The use of solid-state NMR for determining the structures of complex molecules involves MAS decoupling between nuclear spins followed by selective recoupling\cite{Jaroniec-00,Jaroniec-04}.
The recoupling resonances obtained in this paper can, possibly, be used at the latter stage. As discussed in Section~\ref{experiment}, the regime $H_1 \ll H_0$ may provide both the sufficient intensity and the sufficient frequency resolution for such experiments.  If successful, such a method will not require the recoupling sequence to be adjusted to the sample spinning frequency. It will also have an added benefit of the non-trivial spatial dependence in the recoupled Hamiltonians as compared to the usual secular interactions. This dependence can be converted into additional structure information for oriented samples.

In summary, the recoupling resonances for like and unlike spins amount to a new resource for manipulating nuclear spins, and, therefore, are likely to be useful for various NMR applications. However, the condition of applying a high amplitude rf field for a rather long time may be challenging to realize in practice.

\

\section{Conclusion}
\label{conclusion}

In this paper, we described theoretically the properties of recoupling resonances associated with the appearance of non-secular terms in the time-averaged Hamiltonians of nuclear spins-spin interaction in solids in the presence of strong rotating rf fields. 
Our analysis indicates that, under certain conditions, the effect of the recoupled non-secular interaction terms should be particularly well observable. We have also discussed the extension of our findings to the case of linearly polarized rf fields.  Finally, we discussed how the recoupling resonances considered in this paper can used for manipulating nuclear spins in various NMR applications. 

\acknowledgements
We are grateful to J. Haase, J. Kohlrautz, B. Meier and C. P. Slichter for the discussions of this work.

%\appendix

\setcounter{section}{0}
\renewcommand{\thesection}{Appendix A}

\section{Plots of recoupling resonances for unlike spins}
\label{app}

\begin{figure*}[h] \setlength{\unitlength}{0.1cm} 
%=======================================================================

\begin{picture}(170, 200)
{  
\put(21,0){$\gamma_j/\gamma_i$}
\put(61,0){$\gamma_j/\gamma_i$}
\put(-0.5,0){ \epsfxsize= 4cm \epsfbox{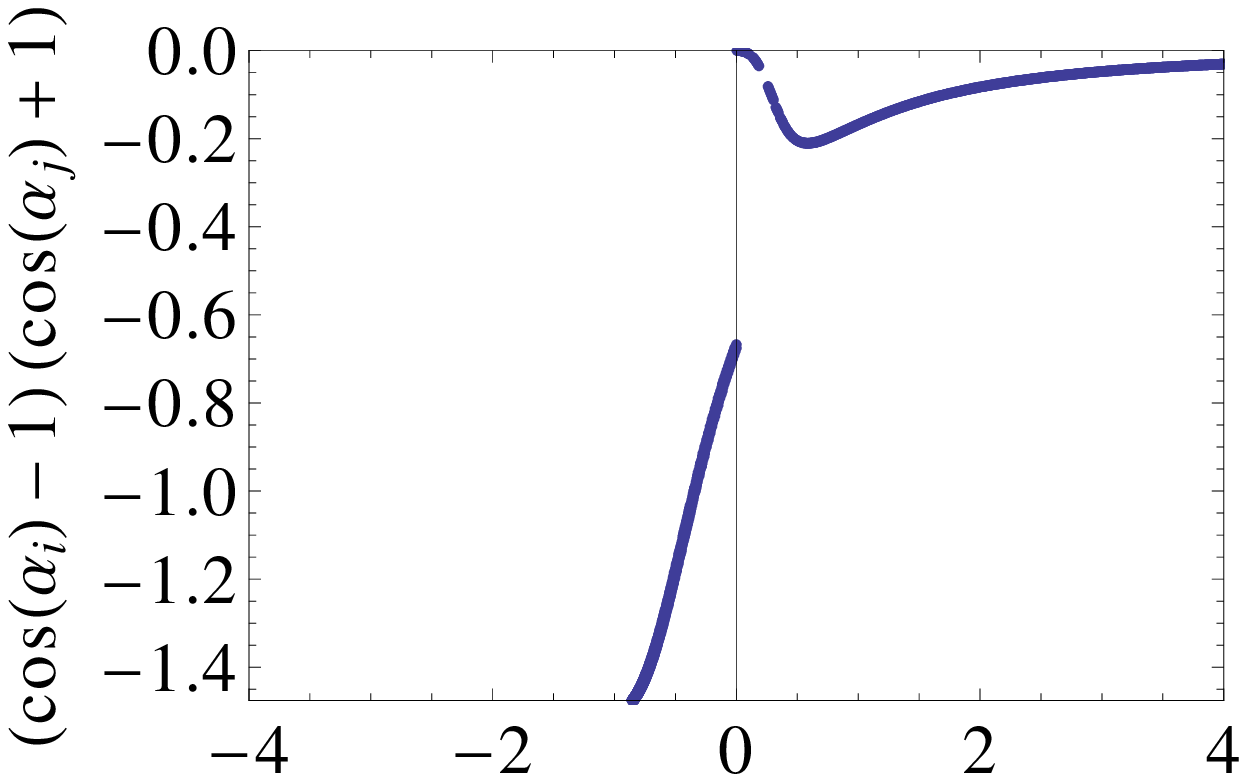} }
\put(41,0){ \epsfxsize= 4cm \epsfbox{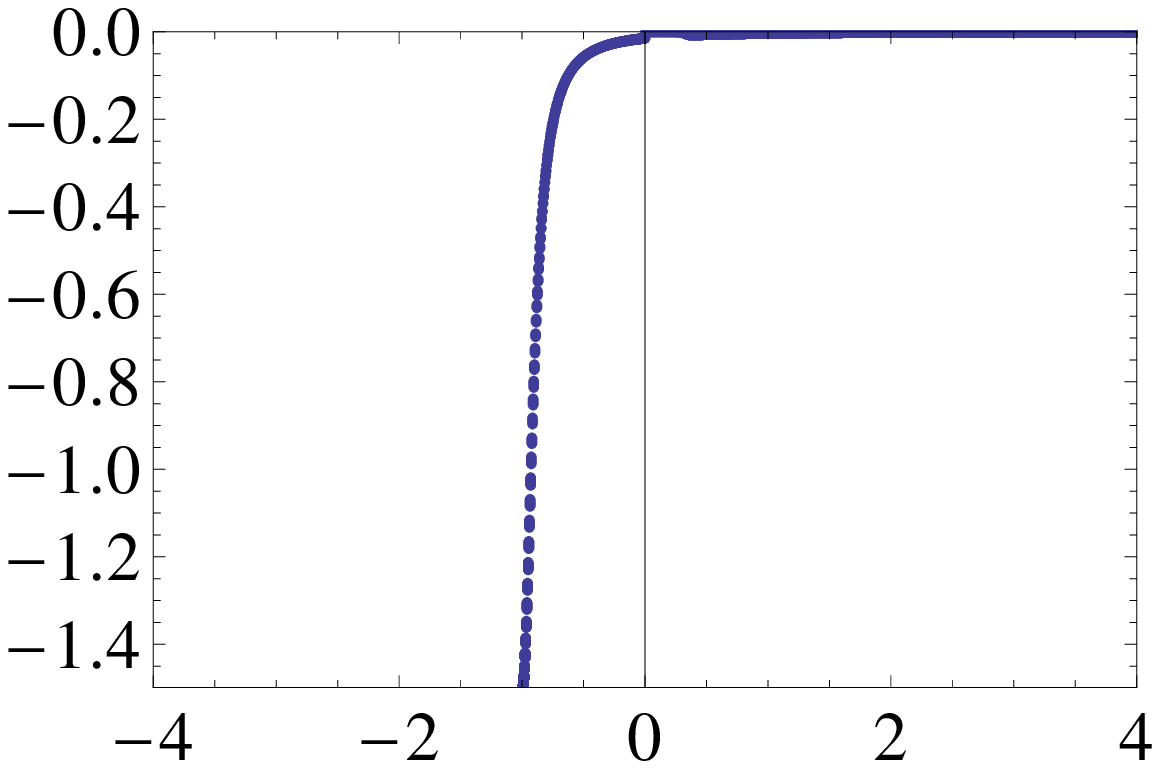} }
\put(0.5,27){ \epsfxsize= 4cm \epsfbox{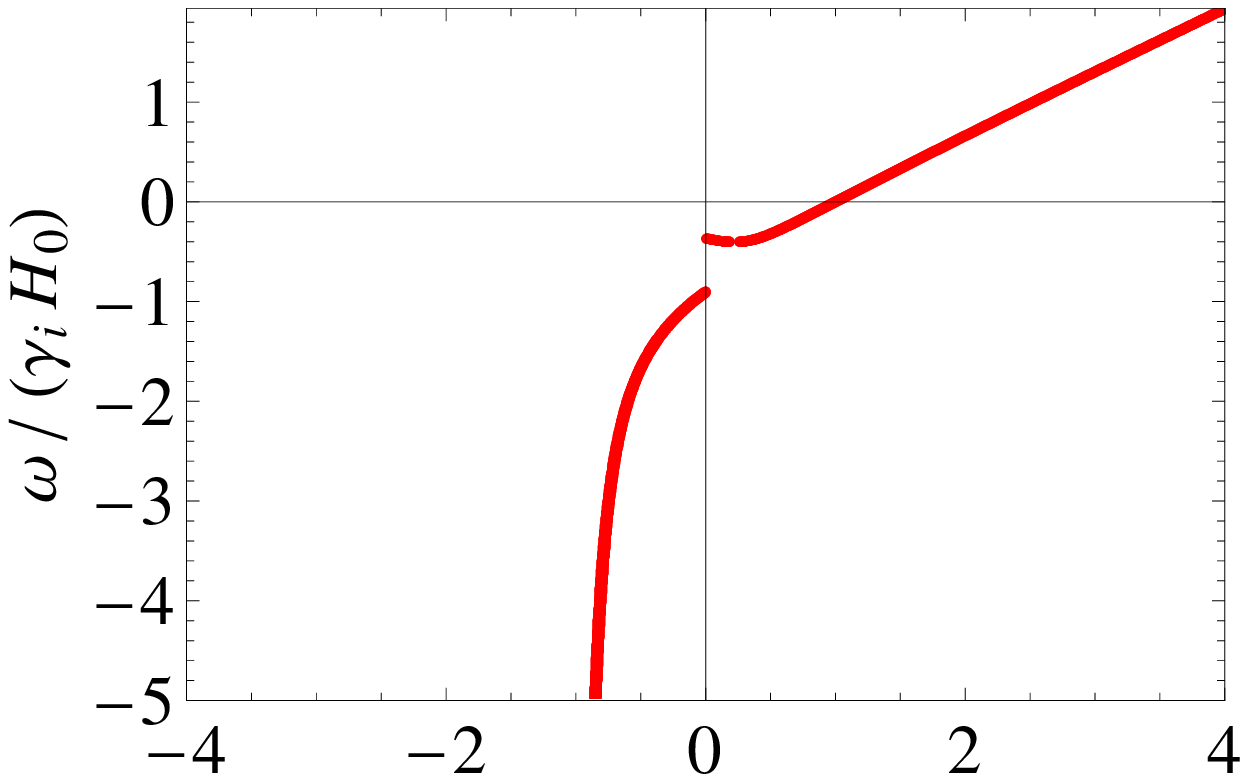} }
\put(41.5,27){ \epsfxsize= 4cm \epsfbox{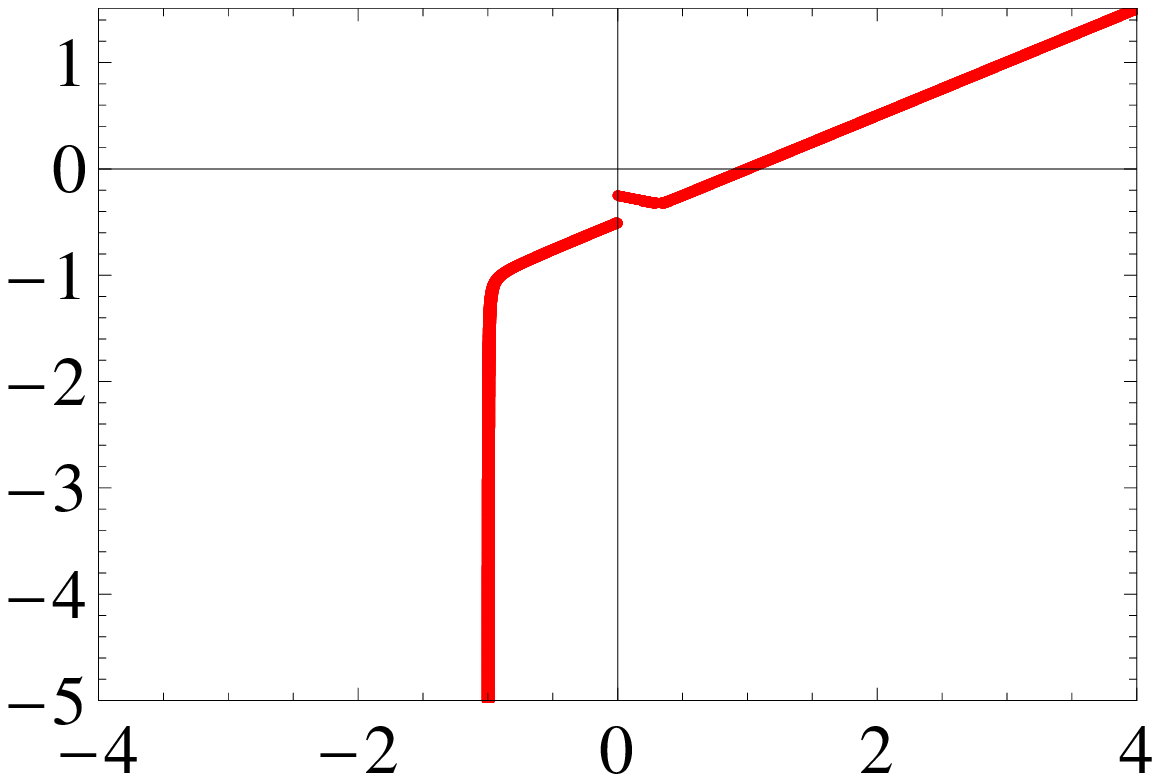} }
\put(33,58){$\omega=-\frac{1}{2}(\Omega_{ei}-\Omega_{ej})$}
\put(0,64){ \epsfxsize= 4cm \epsfbox{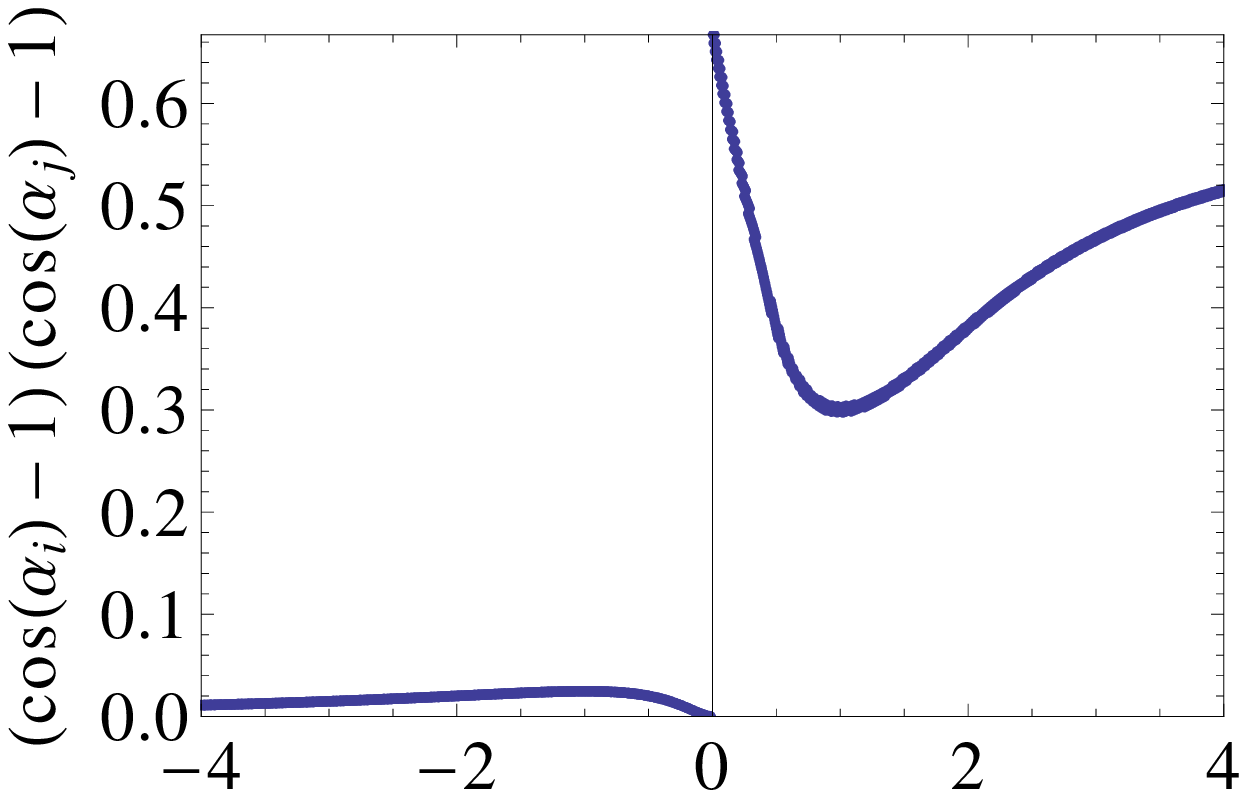} }
\put(40,64){ \epsfxsize= 4cm \epsfbox{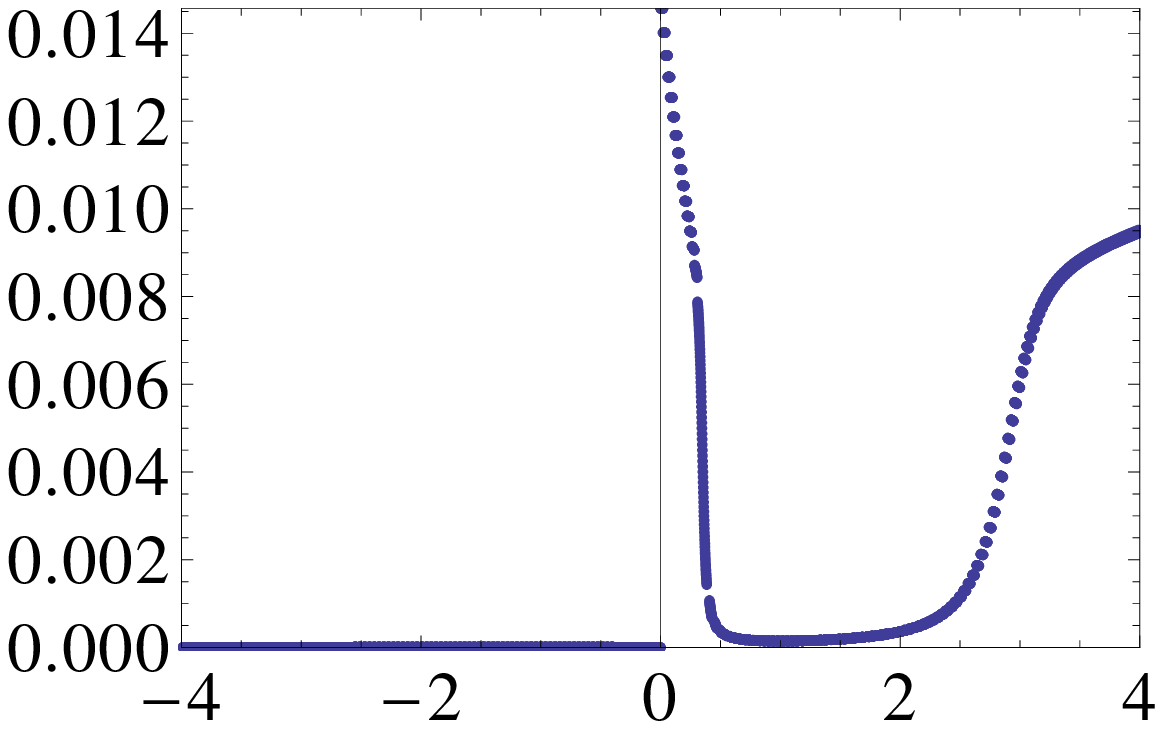} }
\put(0,91){ \epsfxsize= 4cm \epsfbox{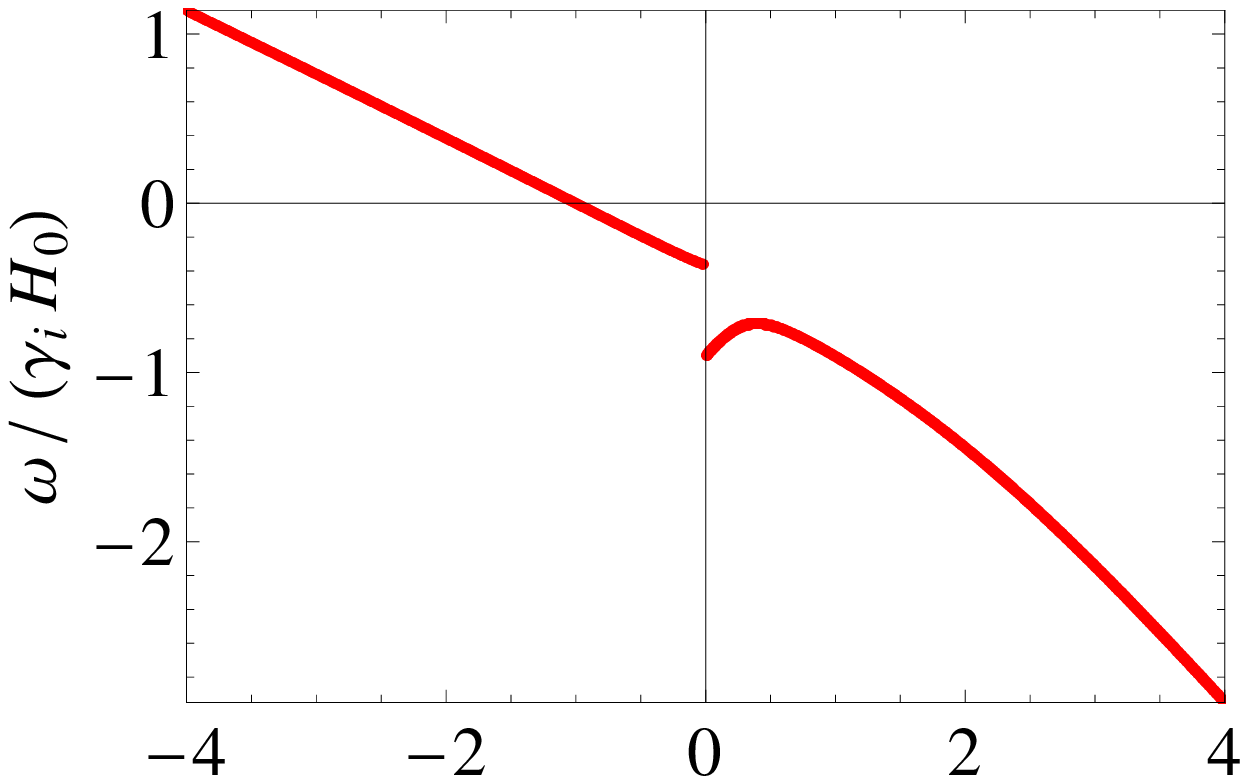} }
\put(40,91){ \epsfxsize= 4cm \epsfbox{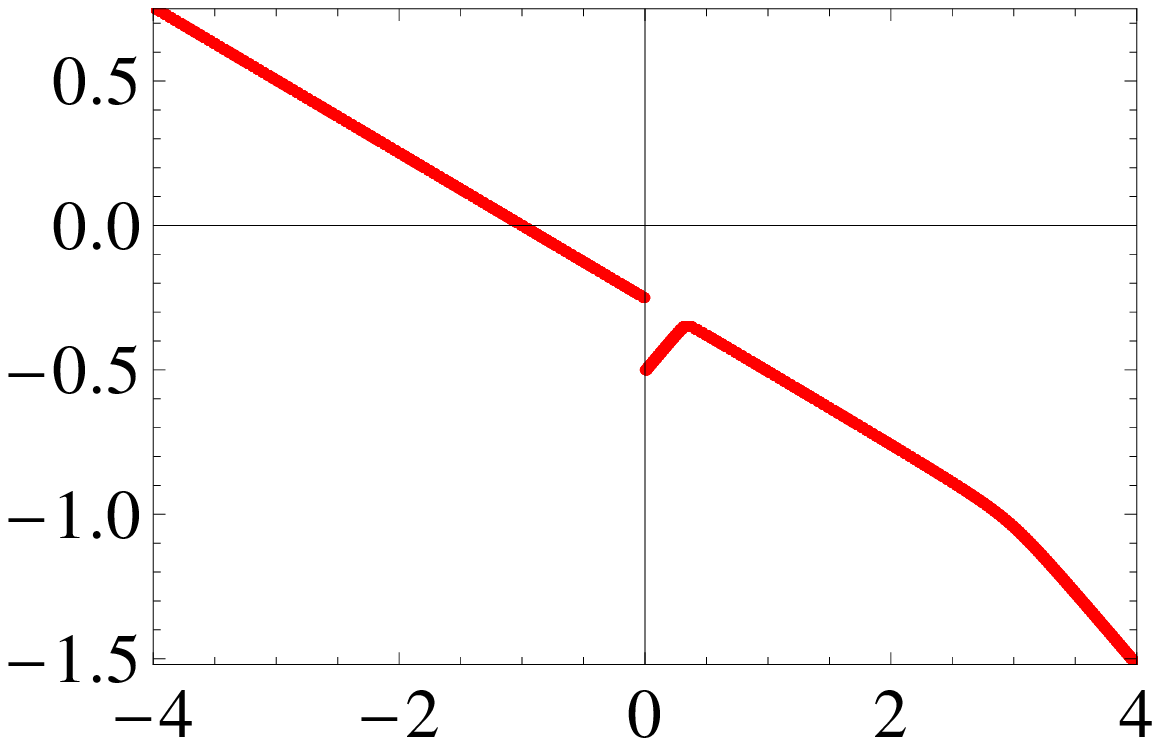} }
\put(33,121){$\omega=-\frac{1}{2}(\Omega_{ei}+\Omega_{ej})$}
\put(-1,127){ \epsfxsize= 4cm \epsfbox{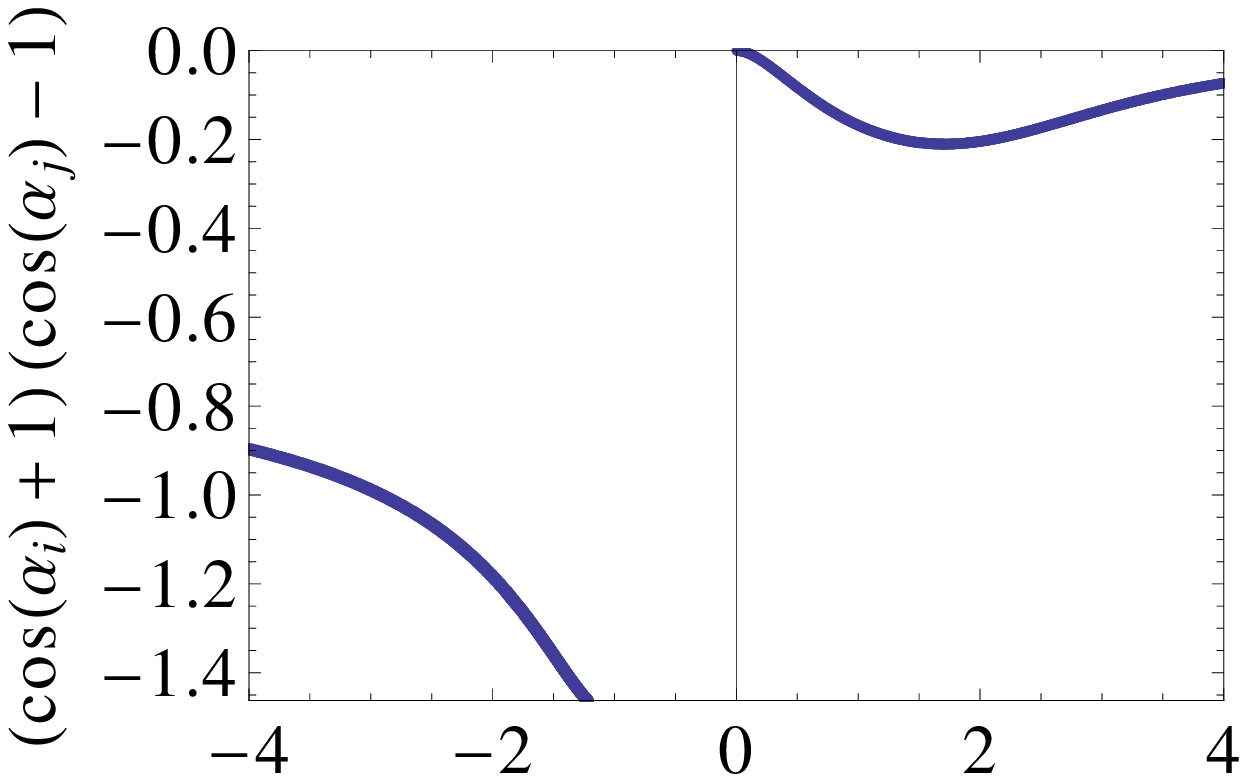} }
\put(39.5,127){ \epsfxsize= 4cm \epsfbox{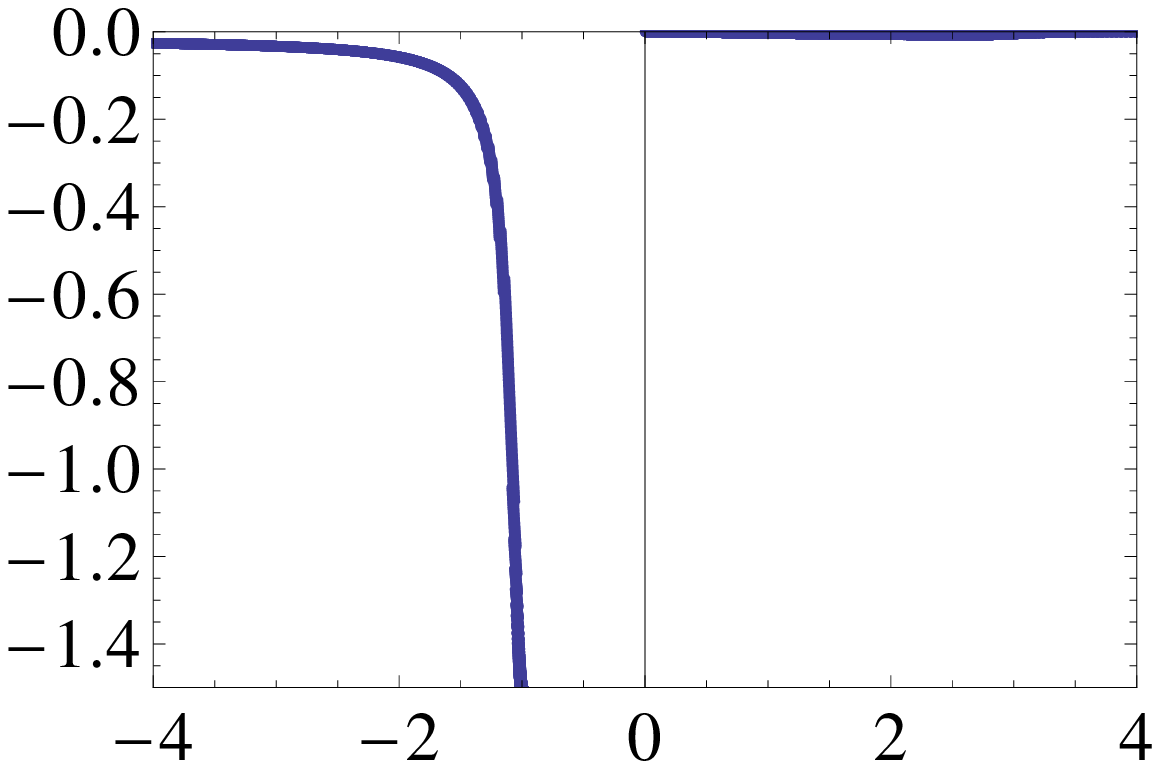} }
\put(0,155){ \epsfxsize= 4cm \epsfbox{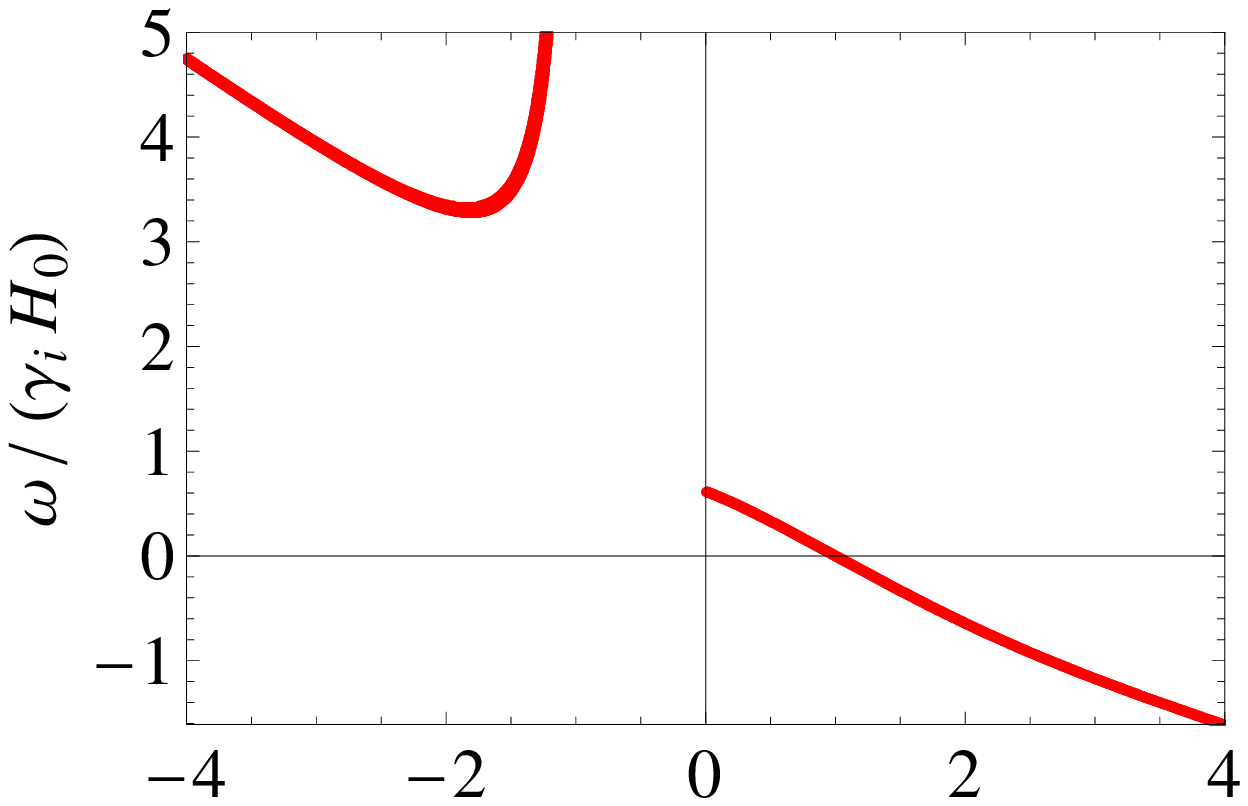} }
\put(40,155){ \epsfxsize= 4cm \epsfbox{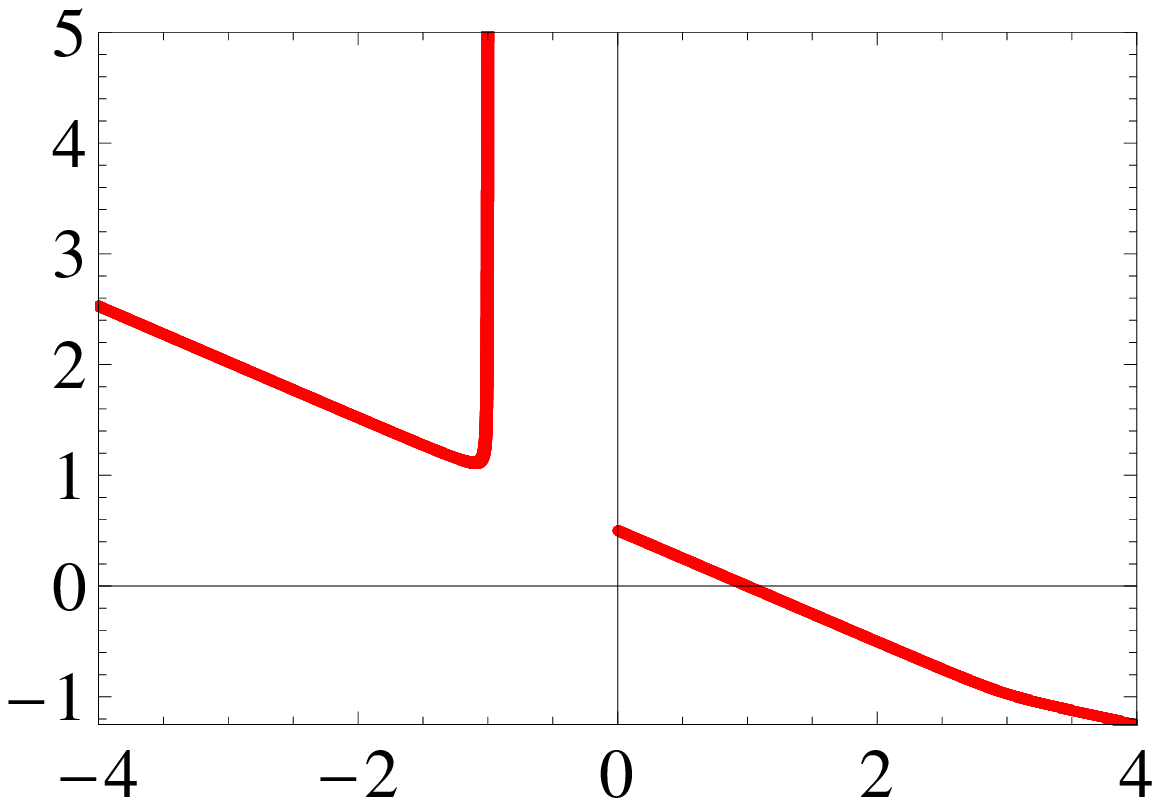} }
\put(33,186){$\omega=\frac{1}{2}(\Omega_{ei}-\Omega_{ej})$}
\put(15,194){$\frac{H_1}{H_0}=0.9$}
\put(55,194){$\frac{H_1}{H_0}=0.1$}
\put(0,187){\bf (a)}
\put(0,122){\bf (c)}
\put(0,59){\bf (e)}

\put(111,0){$\gamma_j/\gamma_i$}
\put(151,0){$\gamma_j/\gamma_i$}
\put(90.5,0){ \epsfxsize= 4cm \epsfbox{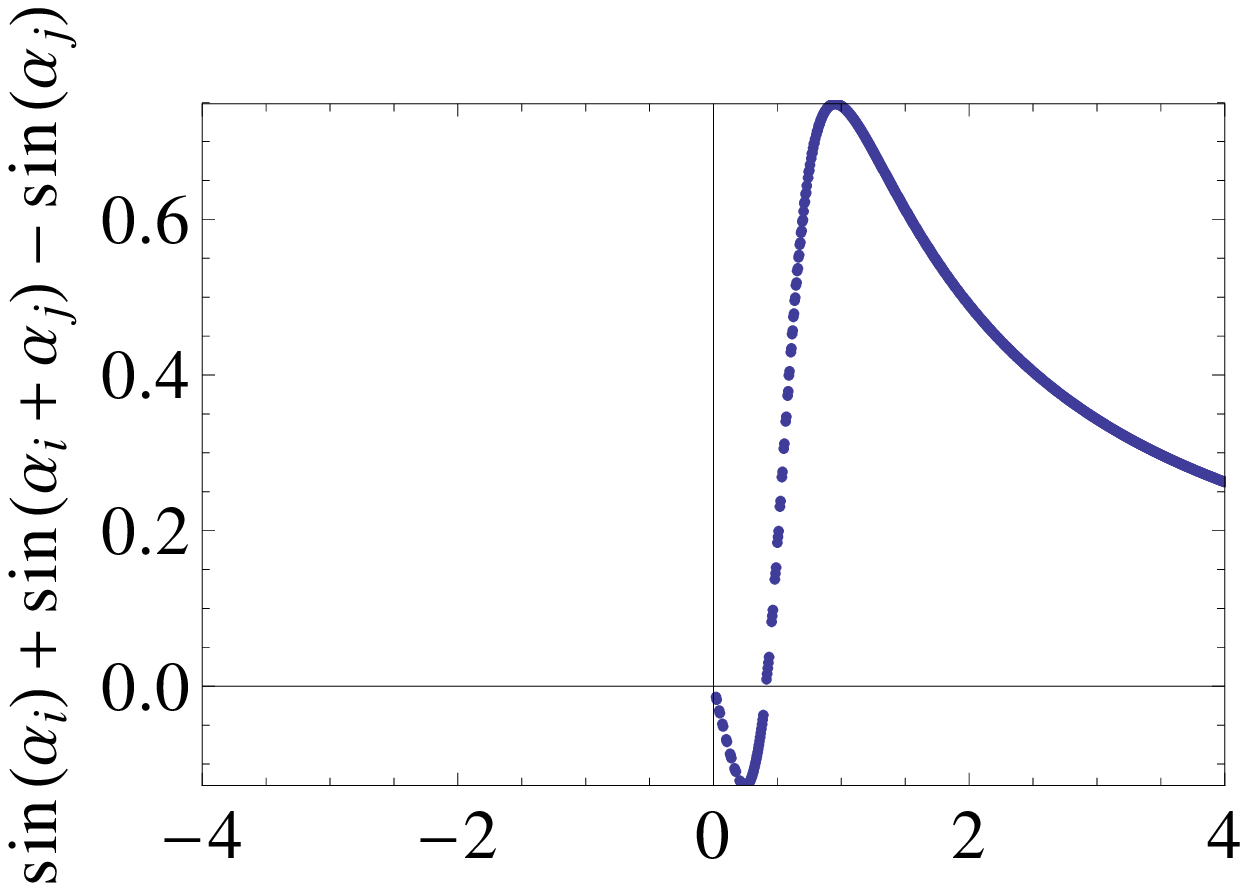} }
\put(130,0){ \epsfxsize= 4cm \epsfbox{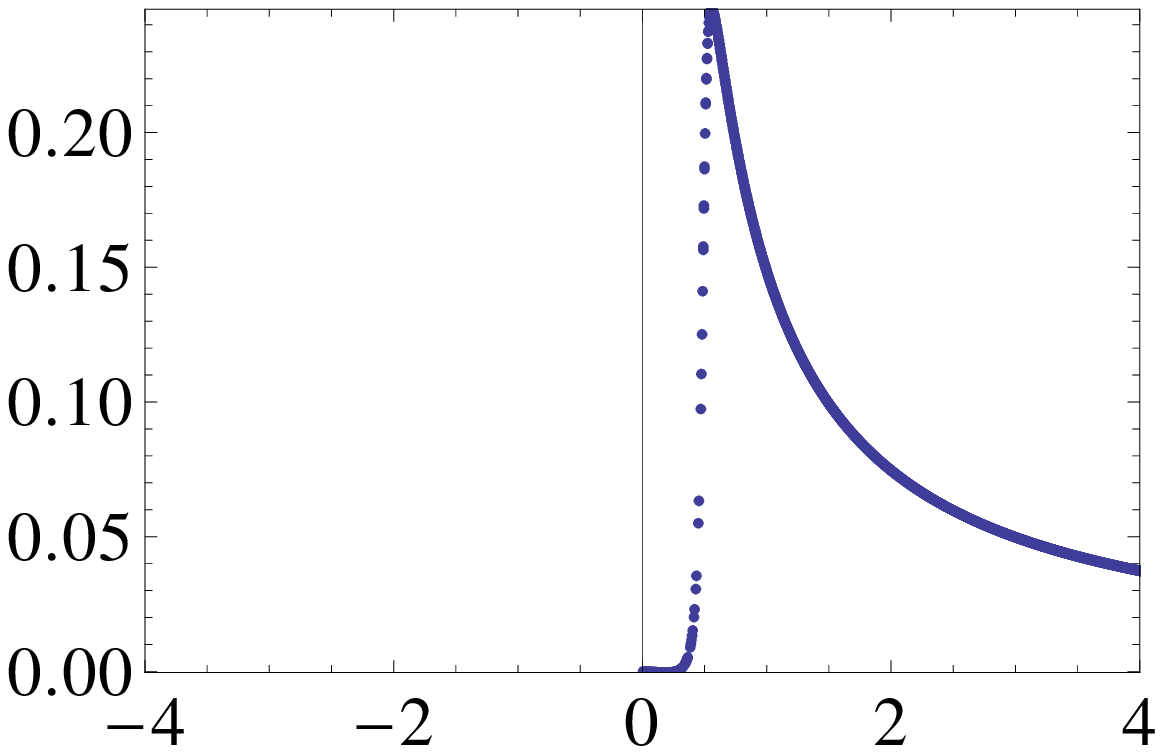} }
\put(92,27){ \epsfxsize= 4cm \epsfbox{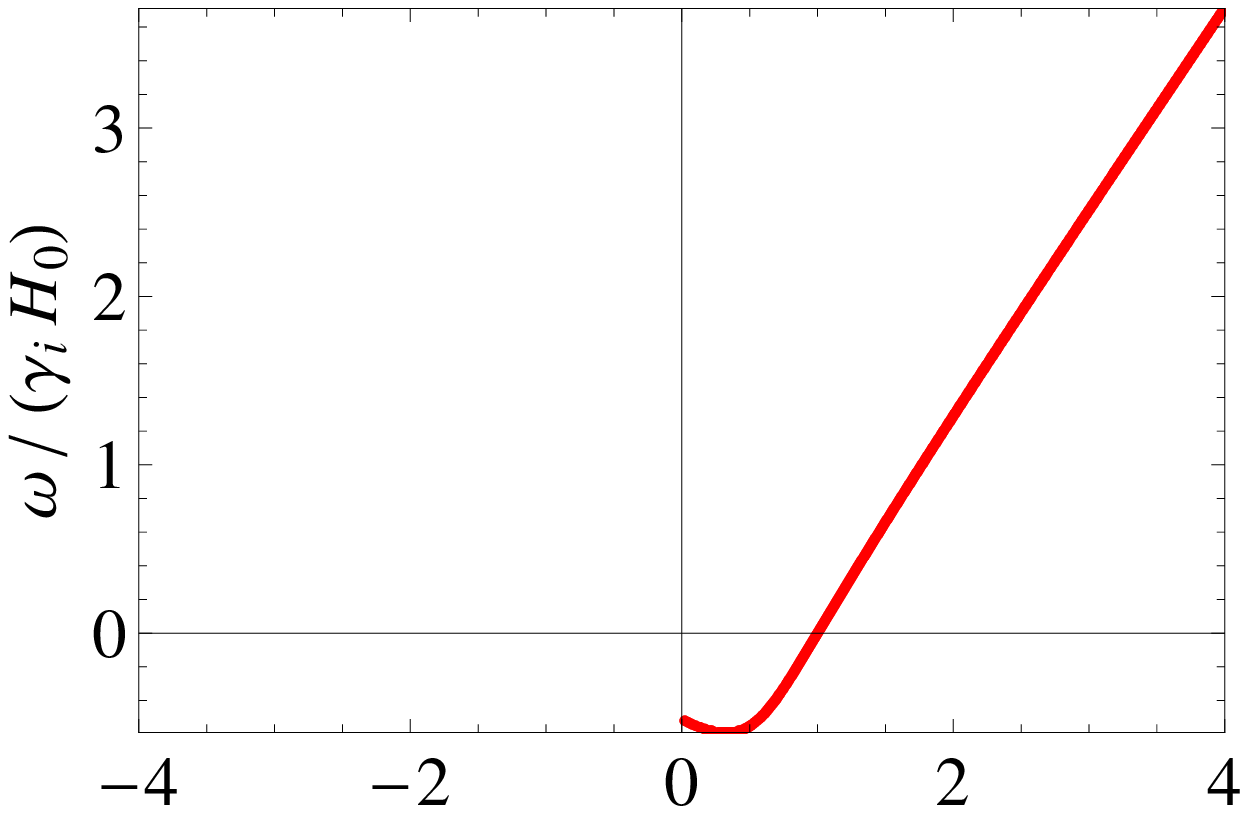} }
\put(131.5,27){ \epsfxsize= 4cm \epsfbox{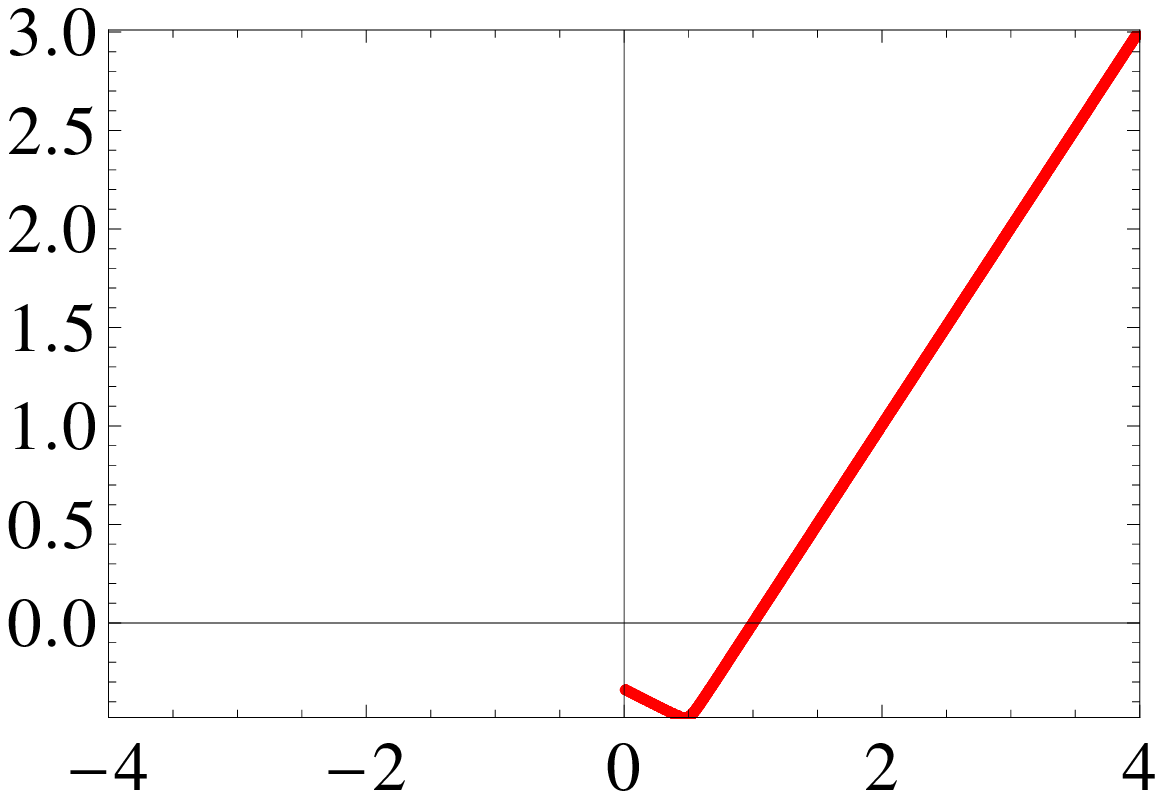} }
\put(123,58){$\omega=-(\Omega_{ei}-\Omega_{ej})$}
\put(89.5,64){ \epsfxsize= 4cm \epsfbox{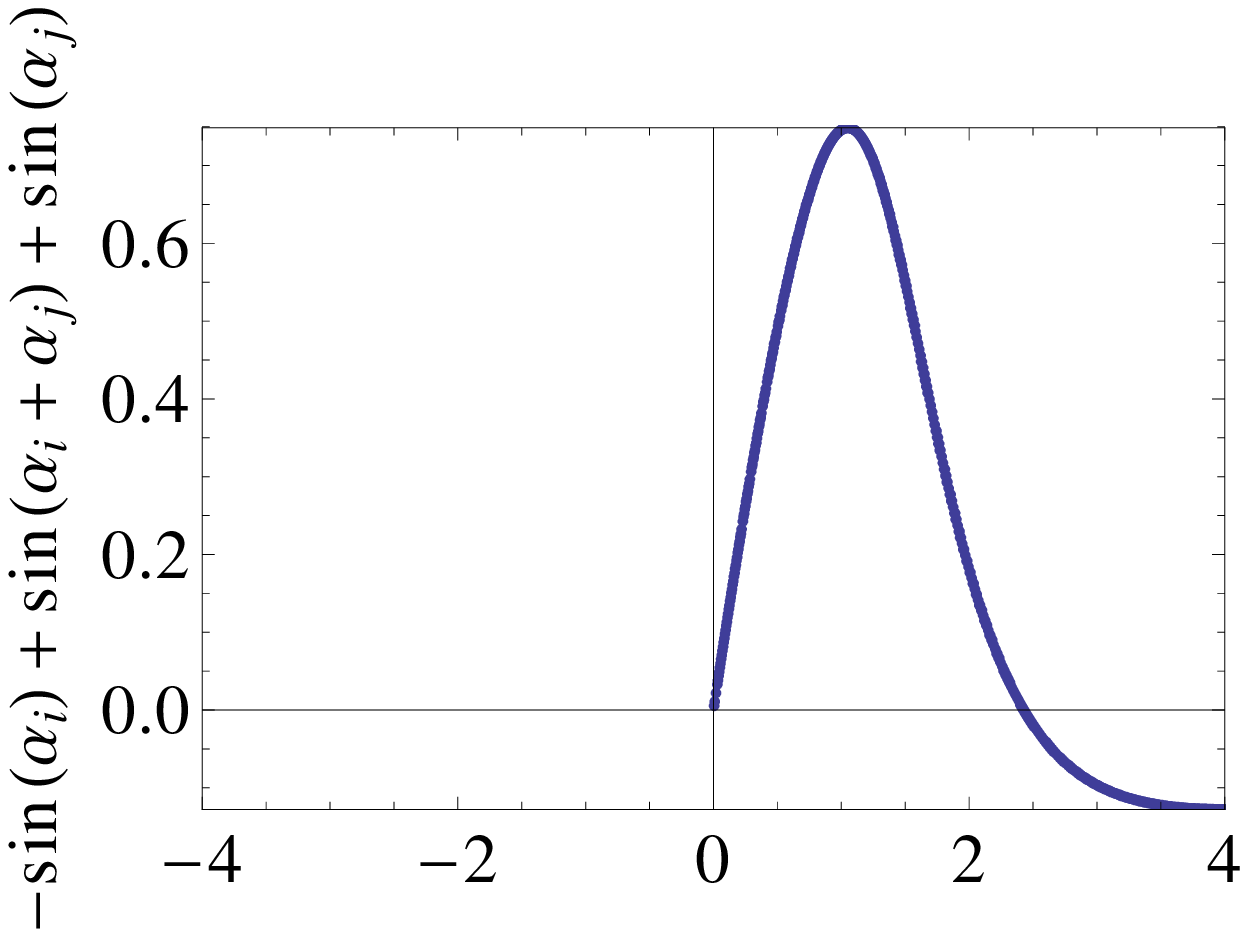} }
\put(129.5,64){ \epsfxsize= 4cm \epsfbox{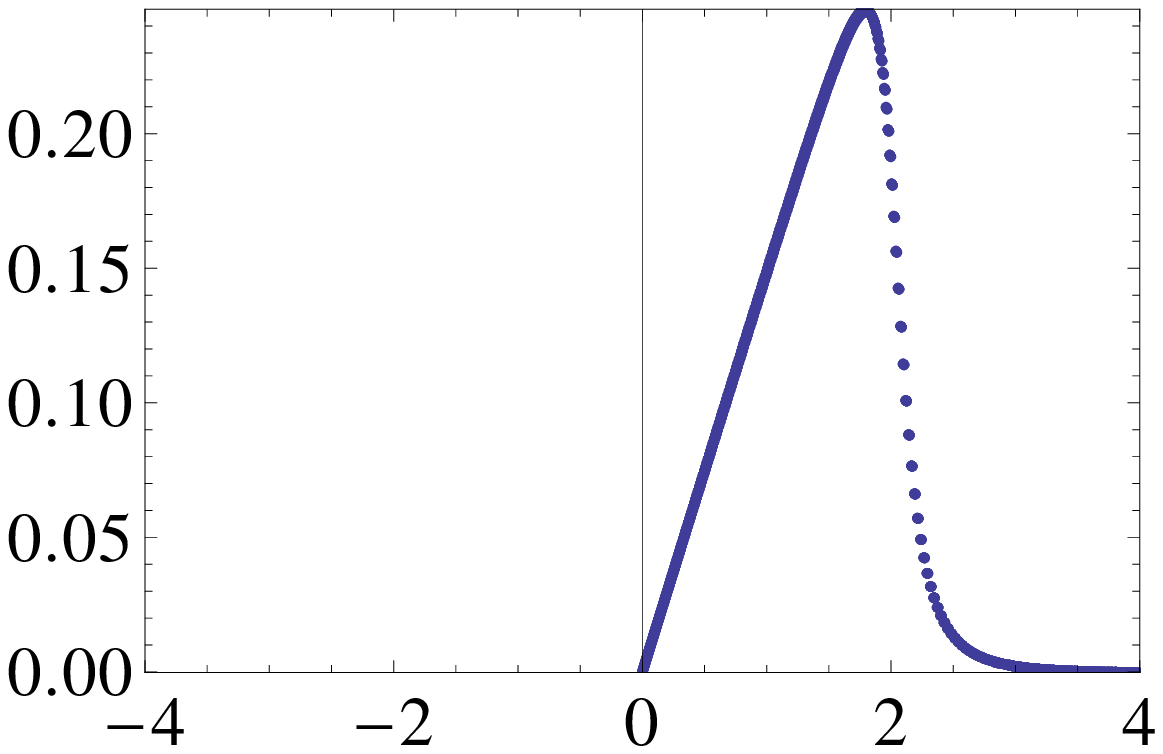} }
\put(90,91){ \epsfxsize= 4cm \epsfbox{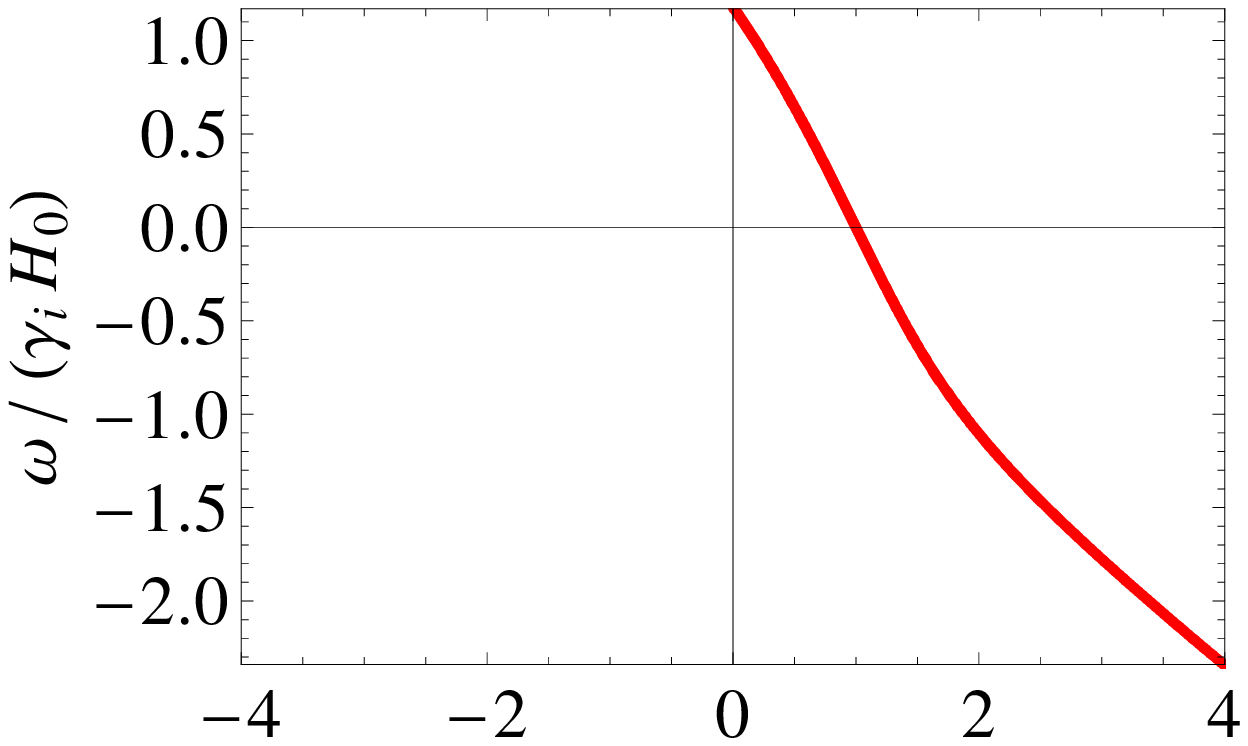} }
\put(130,91){ \epsfxsize= 4cm \epsfbox{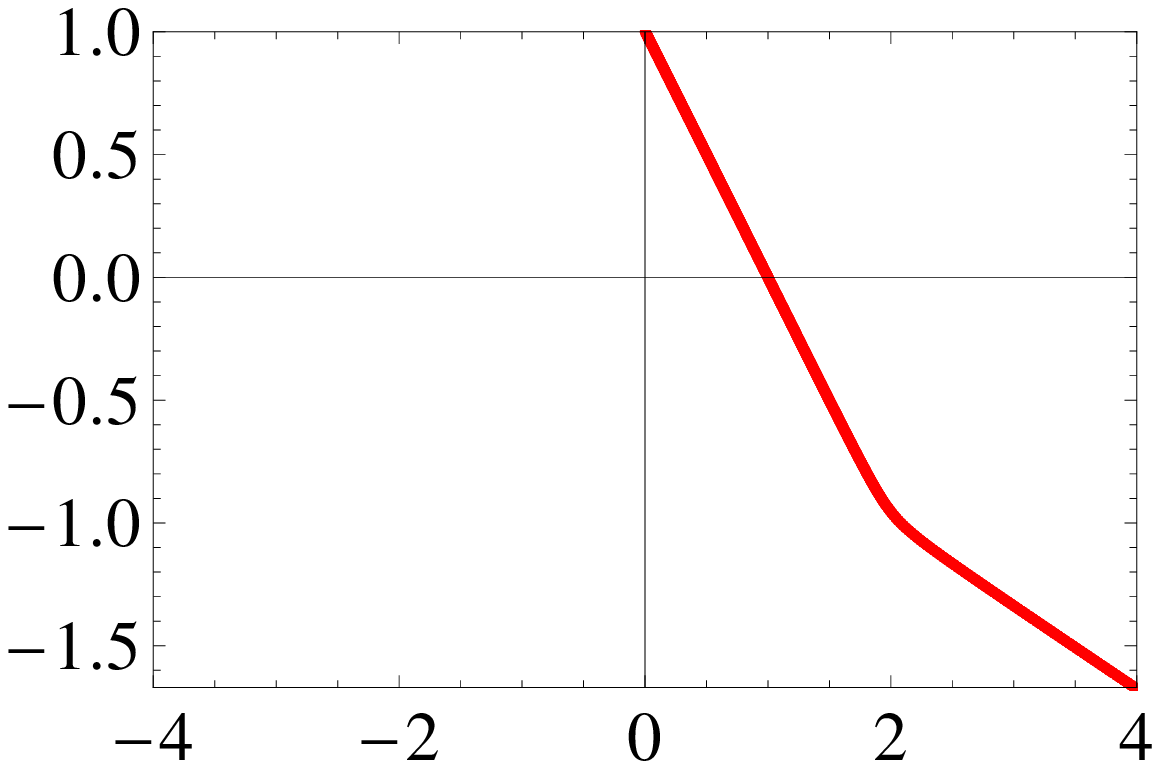} }
\put(123,121){$\omega=(\Omega_{ei}-\Omega_{ej})$}
\put(90,127){ \epsfxsize= 4cm \epsfbox{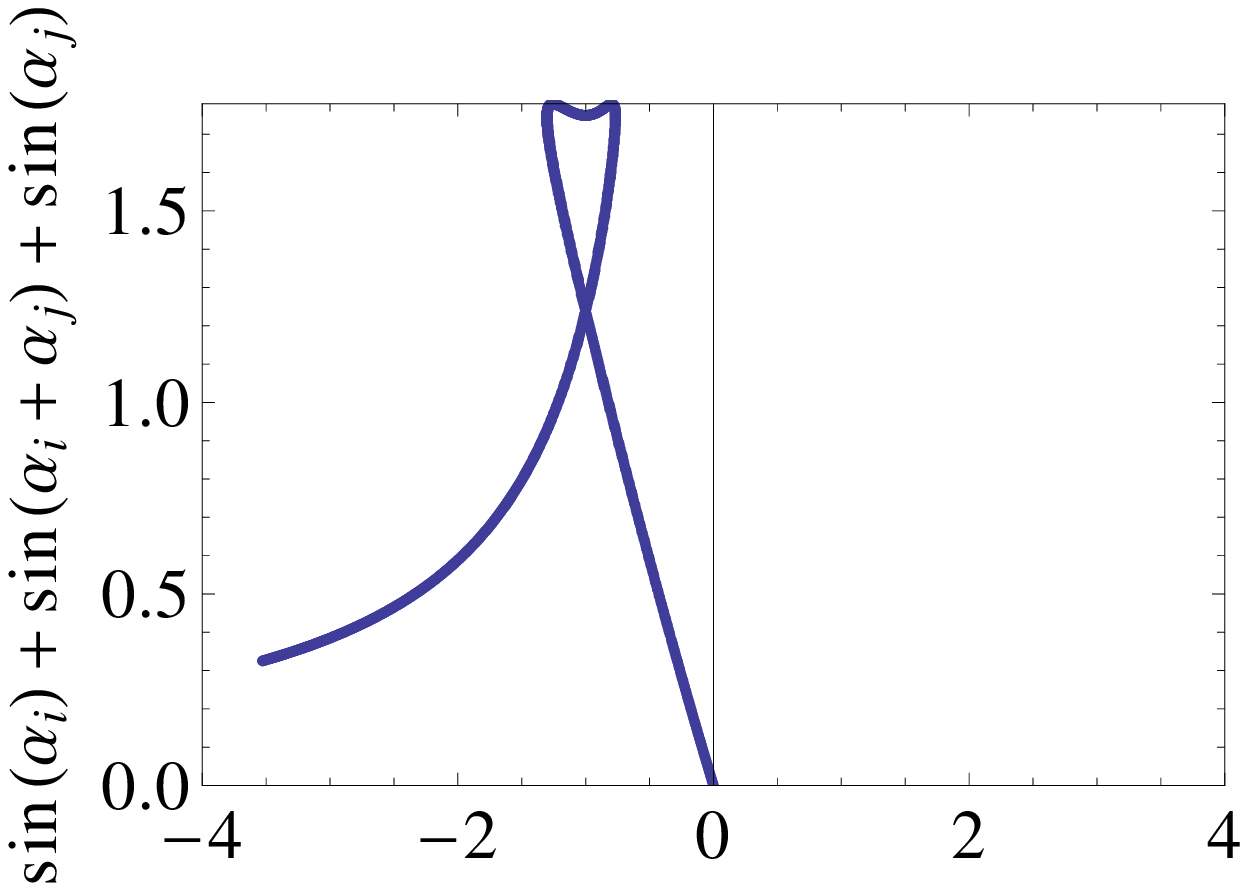} }
\put(130,127){ \epsfxsize= 4cm \epsfbox{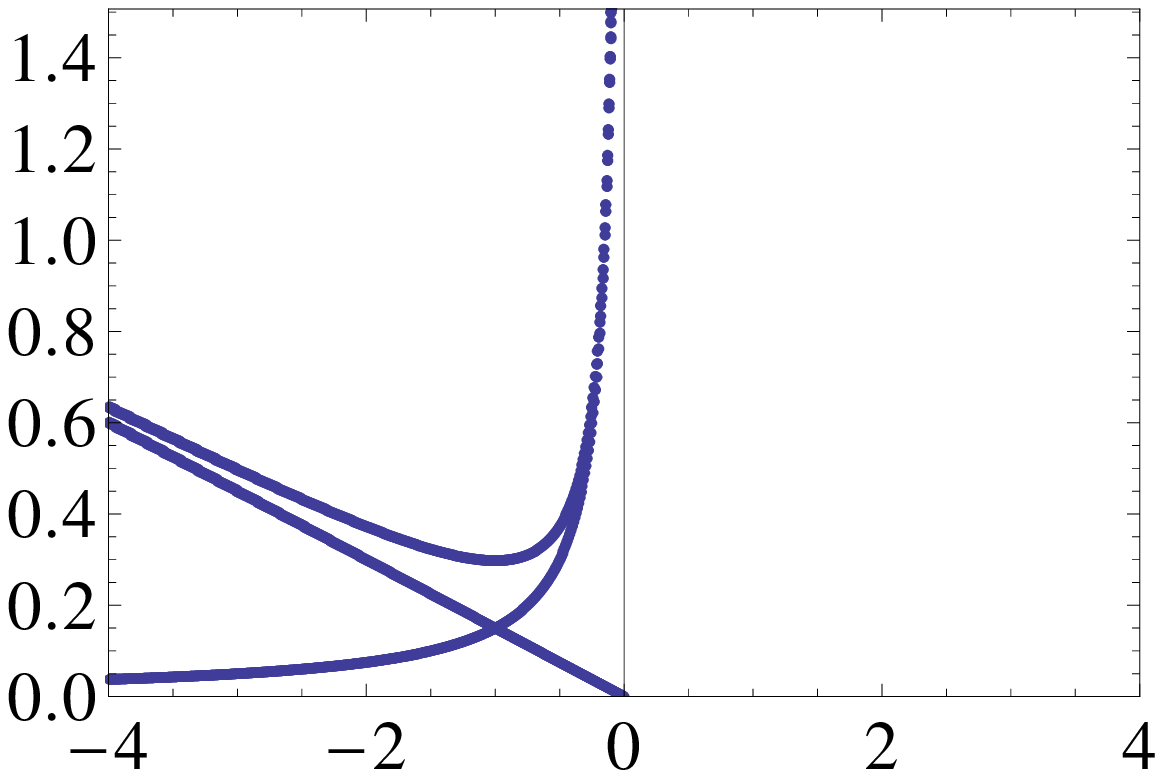} }
\put(90,155){ \epsfxsize= 4cm \epsfbox{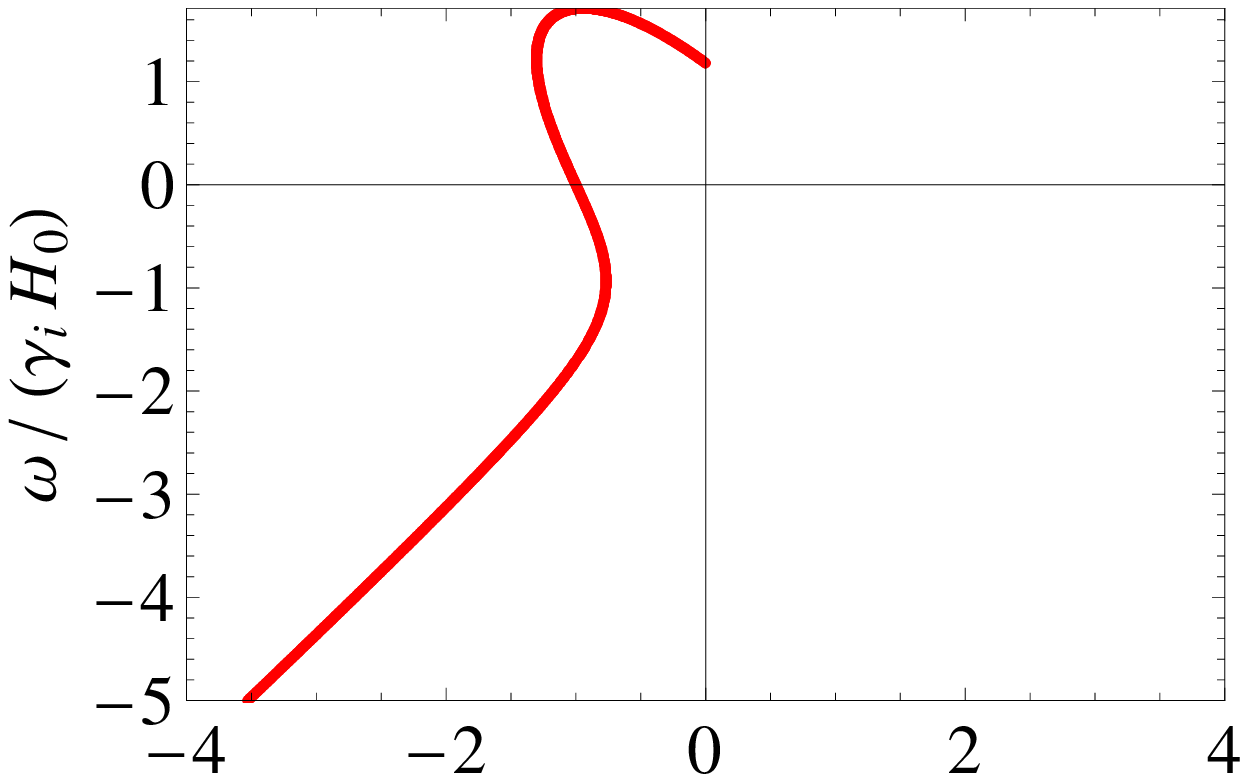} }
\put(130,155){ \epsfxsize= 4cm \epsfbox{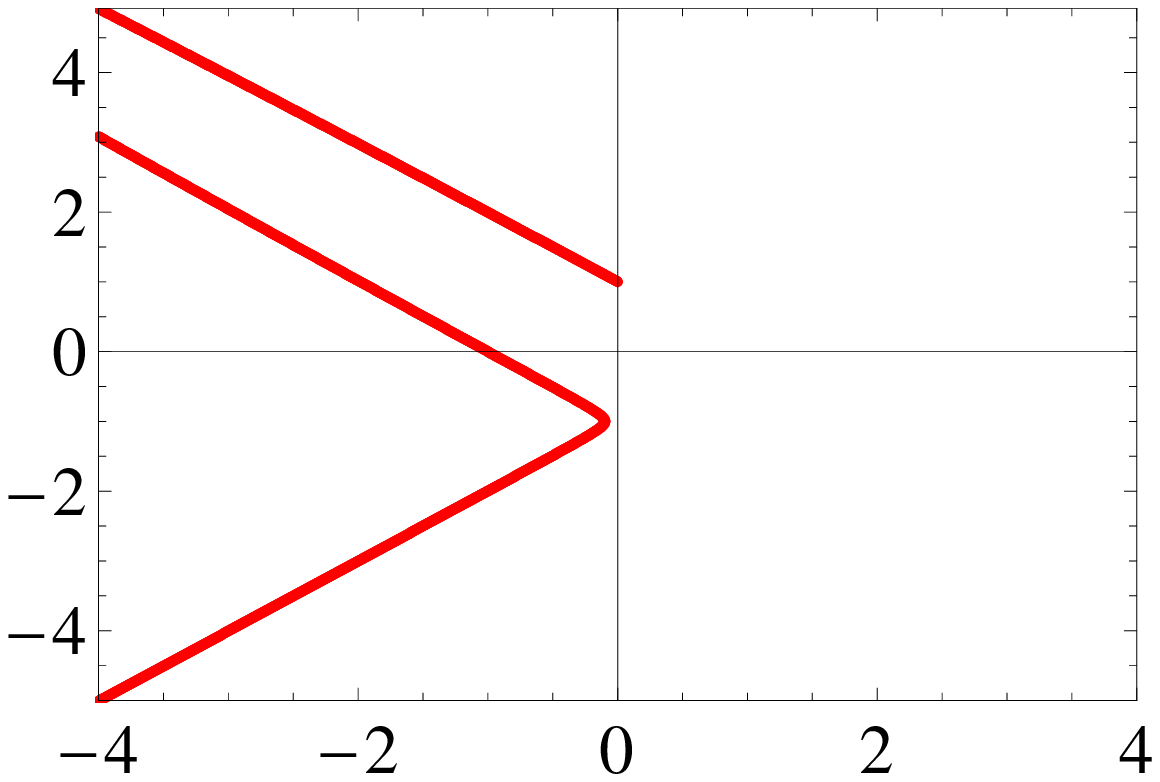} }
\put(123,186){$\omega=(\Omega_{ei}+\Omega_{ej})$}
\put(105,194){$\frac{H_1}{H_0}=0.9$}
\put(145,194){$\frac{H_1}{H_0}=0.1$}
\put(90,187){\bf (b)}
\put(90,122){\bf (d)}
\put(90,59){\bf (f)}
}
\end{picture}
%============== 
\caption{(Color online) Examples of recoupling resonances for unlike spins. Plots of the resonant values of $\omega$ and the prefactors in the recoupled Hamiltonians (\ref{eq:H2dip(om=(Ome1+Ome2))},\ref{eq:H2dip(om=1/2(Ome1+Ome2))}) as functions of $\gamma_j/\gamma_i$  for the resonant conditions indicated above the respective plots. All examples presented are obtained numerically for either $H_1/H_0=0.9$ or $H_1/H_0=0.1$. Each column of plots corresponds to the ratio $H_1/H_0$  indicated above it. The plot lines consist of the actual computed points and thus in some cases have the appearance of dotted lines. 
The results for $H_1/H_0=0.1$ are qualitatively representative of the limit $H_1 \ll H_0$.
The content of this figure together with Fig.~\ref{fig-pref-main} covers all eight resonant conditions associated with Eqs.(\ref{eq:H2dip(om=(Ome1+Ome2))},\ref{eq:H2dip(om=1/2(Ome1+Ome2))})}
\label{fig-pref-app} 
\end{figure*}

This appendix contains the collection of plots (Fig.~\ref{fig-pref-app}) for the six out of eight recoupling resonances described by Eqs.(\ref{eq:H2dip(om=(Ome1+Ome2))},\ref{eq:H2dip(om=1/2(Ome1+Ome2))}) that were not included in Fig.~\ref{fig-pref-main}.

%\begin{widetext}
%%%%%%%%%%%%%%%%%%%%%%%%%%%%%%%%%%%%%%%%%%%%%%%%%%%%%%%%%%%%%%%%%%%
%%%%%%%%%%%%%%%%%%%%%%%%%%%%%%%%%%%%%%%%%%%%%%%%%%%%%%%%%%%%%%%%%%%
%\end{widetext}

%\clearpage

%\bibliography{Kropf_PRB_2012}

\clearpage

\begin{widetext}

\setcounter{figure}{0}
\renewcommand{\thefigure}{S\arabic{figure}}

\setcounter{equation}{0}
\renewcommand{\theequation}{S\arabic{equation}}

\setcounter{section}{0}
\renewcommand{\thesection}{S-\Roman{section}}

\begin{center}

{\bf \large SUPPLEMENTARY MATERIAL }

\end{center}

\

\

Note 1: In this supplement, we use the same notations as in the main article unless explicitly specified otherwise. 

\

Note 2: Some of the matrices in this supplement are computer-generated and presented as figures rather than equations. 

\section{Formalism for the calculations of time-averaged spin-spin interaction Hamiltonians}

We denote the full spin-spin interaction Hamiltonian in the double-rotated reference frame as $\Hdip' (t)$. Our goal below is, first, to obtain the explicit expression for $\Hdip' (t)$ and then to average it over time according to formula
\begin{equation}
      \langle \Hdip' \rangle = \lim_{\T \to \infty} \frac{1}{2\T}\int_{-\T}^{\T} \Hdip' (t)dt.
\label{eq:Hm}
\end{equation}

The Hamiltonian $\Hdip' (t)$ can be represented in tensor notations as follows
\begin{equation}
	\Hdip' (t)=\sum_{i<j}^N \J_{ij}\cdot{I}'_{i,\mu} \;\A(ij)_{\mu\nu}\;{I}'_{j,\nu},
	\label{eq:Hdip(t) an}
\end{equation}
where we use the Einstein convention for summing over repeated Greek-letter matrix and vector indices, which can take values $1,2,3$ and correspond, respectively, to the $x,y,z$ projections in the relevant reference frame. The matrix $\A(ij)$ is given by the following equation:
\begin{equation}
	\A(ij) = P_{\mu\delta}(i)C_{\delta\rho}(ij)Q_{\rho\nu}(j),
	\label{eq:A(ij) an} 
\end{equation}
where time-independent matrix $\Cij(ij)$ represents the interaction coefficients of the full interaction Hamiltonian $\Hdip$ in the laboratory reference frame, while time-dependent matrices $\Pij(i)$ and $\Q(j)$ represent the transformations between the laboratory and the double-rotated reference frames appropriate for spins $i$ or $j$, respectively. Matrix $\Cij(ij)$ is presented in Fig.~\ref{fig:C(ij) an}. 
\begin{figure}[H]
	\centering
	\includegraphics[width=0.5\textwidth]{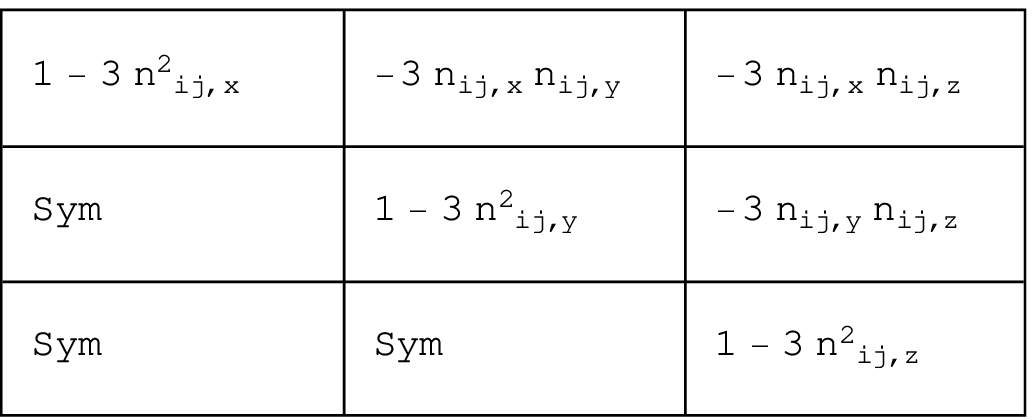}
	\caption{Explicit form of the matrix $\Cij(ij)$. The factors $\n_{ij,\mu}=\frac{{r}_{ij,\mu}}{\arrowvert{{r}_{ij}}\arrowvert}$ are time-independent material constants. The matrix $\Cij(ij)$ is symmetric.}
	\label{fig:C(ij) an}
\end{figure}

The expressions for the transformation matrices are:
\begin{equation}
\Pij(i) = \U{}{\Ome[i],\mu\delta}U_{\alpha_i,\delta\rho}U_{\Om,\rho\nu},
\label{Pij}
\end{equation}
and
\begin{equation}
\Q(j)_{\mu\nu} =  U^{T}_{\Om,\mu\delta}U^{T}_{\alpha_j\delta\rho}U^{T}_{\Ome[j],\rho\nu},
\label{Qj}
\end{equation}
where
\begin{equation}
	\U{}{\Om}=\begin{pmatrix} \cos(\Om t) & \sin (\Om t) & 0 \\ -\sin(\Om t) & \cos(\Om t) & 0 \\ 0 & 0 & 1\end{pmatrix},
	\label{eq:V(Om) an}
\end{equation}
\begin{equation}
	\U{}{\alpha_{i}}=\begin{pmatrix} \cos (\alpha_{i}) & 0 & -\sin (\alpha_{i}) \\ 0 & 1 & 0\\ \sin (\alpha_{i})  & 0 & \cos (\alpha_{i}) \end{pmatrix},
	\label{eq:V(alpha) an}
\end{equation}
\begin{equation}
	\U{}{\Ome[i]}=\begin{pmatrix} \cos(\Ome[i]\cdot  t) & -\sin(\Ome[i] \cdot t) & 0 \\ \sin(\Ome[i]\cdot t) & \cos(\Ome[i,j]\cdot  t) & 0 \\ 0 & 0 & 1\end{pmatrix}
	\label{eq:V(Ome) an}.
\end{equation}
These are the same matrices as those defined by Eqs.(7-9) of the main article but with the spin and matrix indices explicitly included. The explicit form of matrices
$\Pij(i)$ and $\Q(j)$ is given in Figs.~\ref{fig:P(i) an} and \ref{fig:Q(j) an}, respectively.

\begin{figure}[H]
	\centering
	\includegraphics[width=\textwidth]{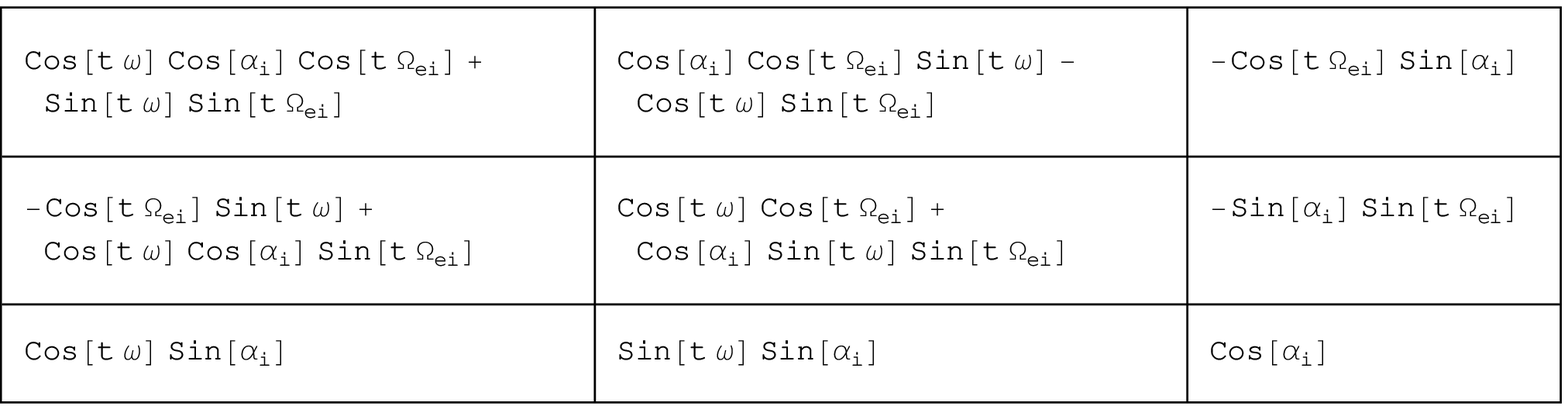}
	\caption{Explicit form of the matrix $\Pij(i)$}
	\label{fig:P(i) an}
\end{figure}
\begin{figure}[H]
	\centering
	\includegraphics[width=\textwidth]{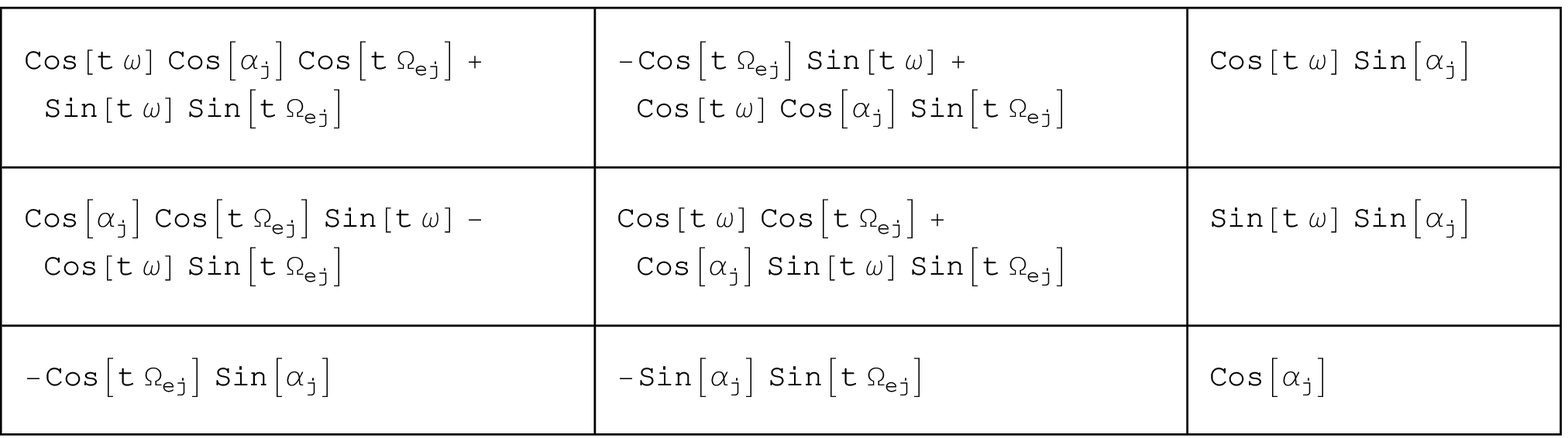}
	\caption{Explicit form of the matrix $\Q(j)$}
	\label{fig:Q(j) an}
\end{figure}
%\pagebreak

The calculation of the time-average of $\Hdip' (t)$ given by Eq.(\ref{eq:Hm}) requires calculating 
\begin{equation}
	 \langle \A(ij) \rangle = \lim_{\T \to \infty} \frac{1}{2\T}\int_{-\T}^{\T} \A(ij) (t)dt
	 \label{eq:A Av an}
\end{equation}
The full and time-averaged expressions for matrix $ \A(ij) (t) $ are presented, for the case of like spins, in Section \ref{sec:Results like an} and, for the case of unlike spins, in Section \ref{sec:Results unlike an}. 

To describe the different recoupling resonances we use ``normalized delta-function'' symbol $\delta(x)$ defined as follows:
\begin{equation}
	\delta(x)=
	\begin{cases}
		1 \;&\text{ if }\; x=0;\\
		0 \;&\text{ otherwise.}
	\end{cases}
	\label{eq:Delta an}
\end{equation}

\section{Like spins}\label{sec:Results like an}
For like spins, we use the notation
\begin{equation}
	 \alpha_i=\alpha_j\equiv\alpha
\end{equation}
and
\begin{equation}
	\Ome[i]=\Ome[j]\equiv\Ome.
\end{equation}

In this case, $\Pij(i)=\Q(j)$, and, as a result, $\A(ij)$ is symmetric. The time-dependent matrix $\A(ij)$ is presented in function of the elements of $\Cij(ij),\;\Pij(i)$ and $\Q(j)$ in Fig.~\ref{fig:Alike PQ an}. The explicit form of $\A(ij)$ as a function of $\alpha$, $\Om$ and $\Ome$ is given in Fig.~\ref{fig:Alike Om an}. The time-averaged value of $\A(ij)$ is presented in Fig.~\ref{fig:Alike Av an}. 

\begin{figure}[H]
	\centering
	\includegraphics[width=\textwidth]{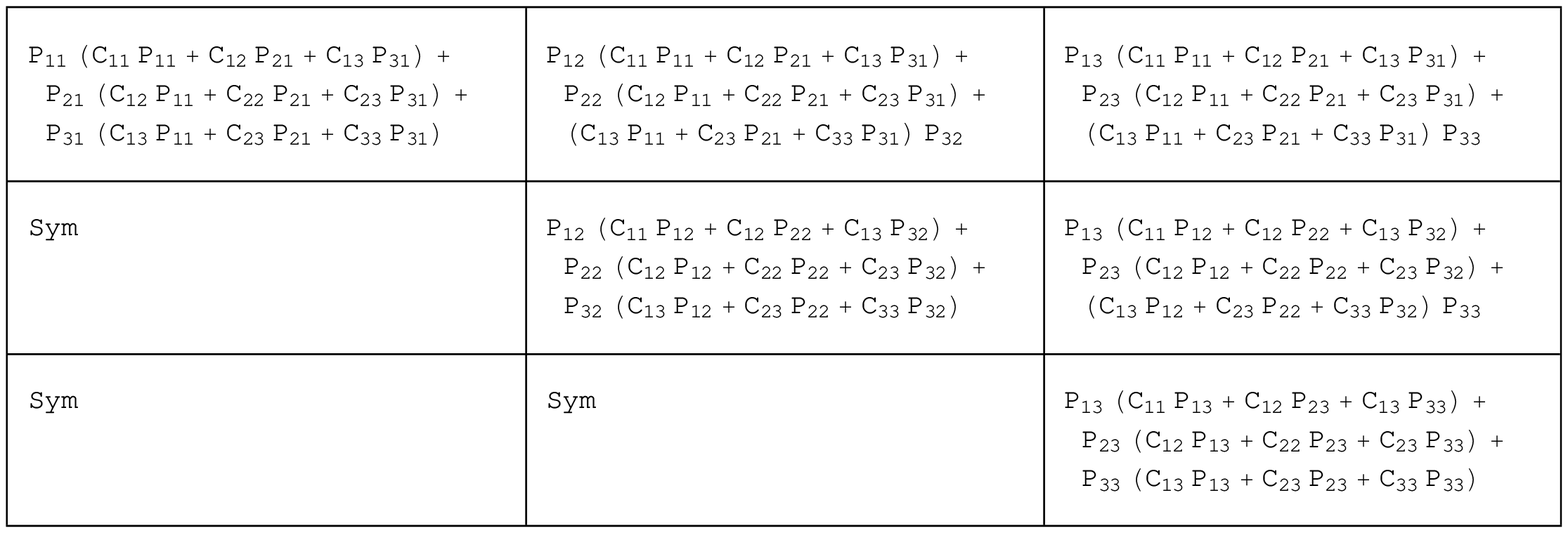}
	\caption{Like spins : The matrix $\A(ij)$ in compact form. Here $C_{\mu\nu} \equiv \Cij(ij)$, $P_{\mu\nu} \equiv \Pij(i)$ and $Q_{\mu\nu} \equiv \Q(j)$. }
	\label{fig:Alike PQ an}
\end{figure}
\begin{figure}[H]
	\centering
	\includegraphics[height=0.6\textheight,width=\textwidth]{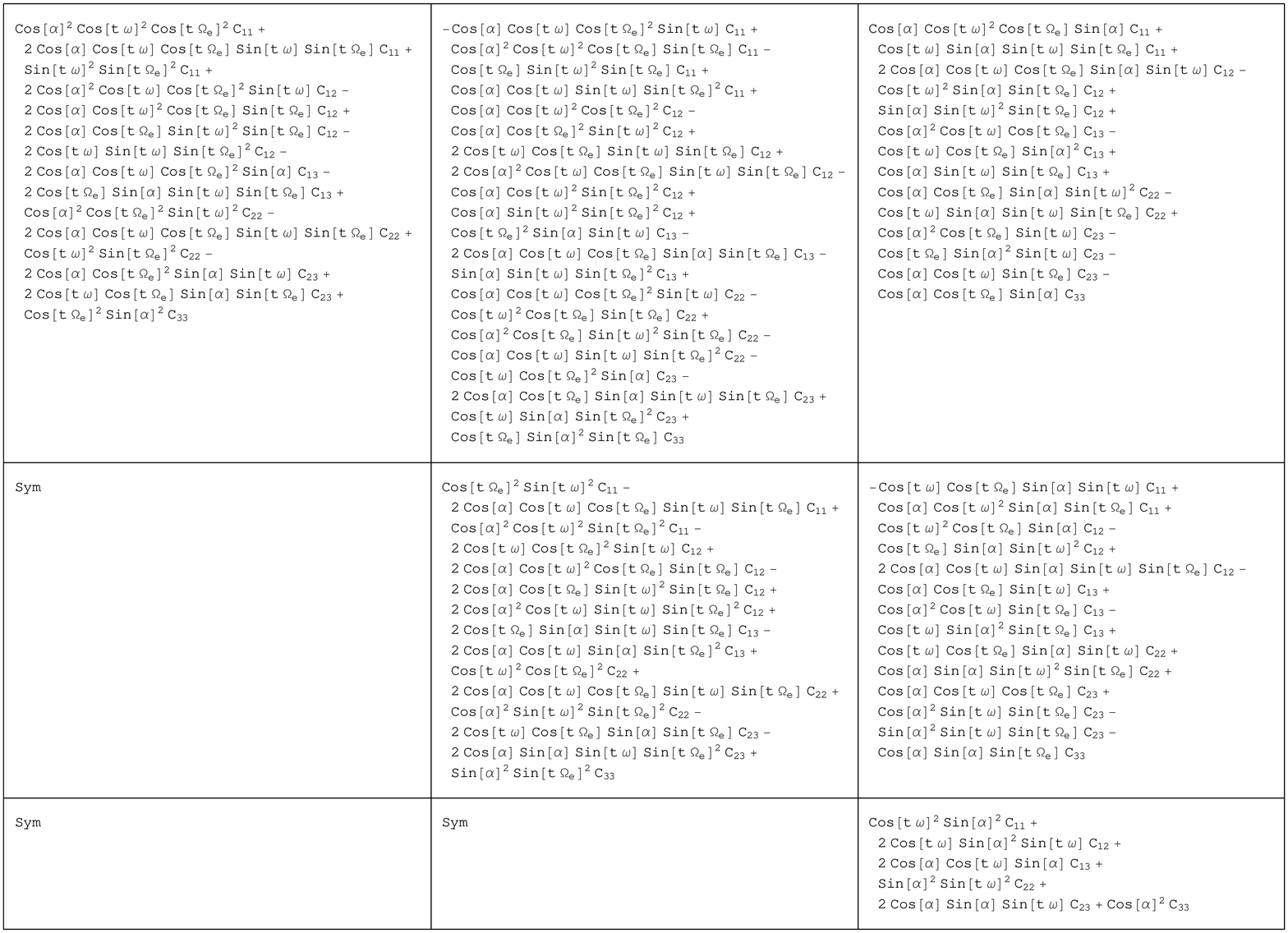}
	\caption{Like spins : Explicit form of the matrix $\A(ij)$}
	\label{fig:Alike Om an}
\end{figure}
\begin{figure}[H]
	\centering
	\includegraphics[width=\textwidth]{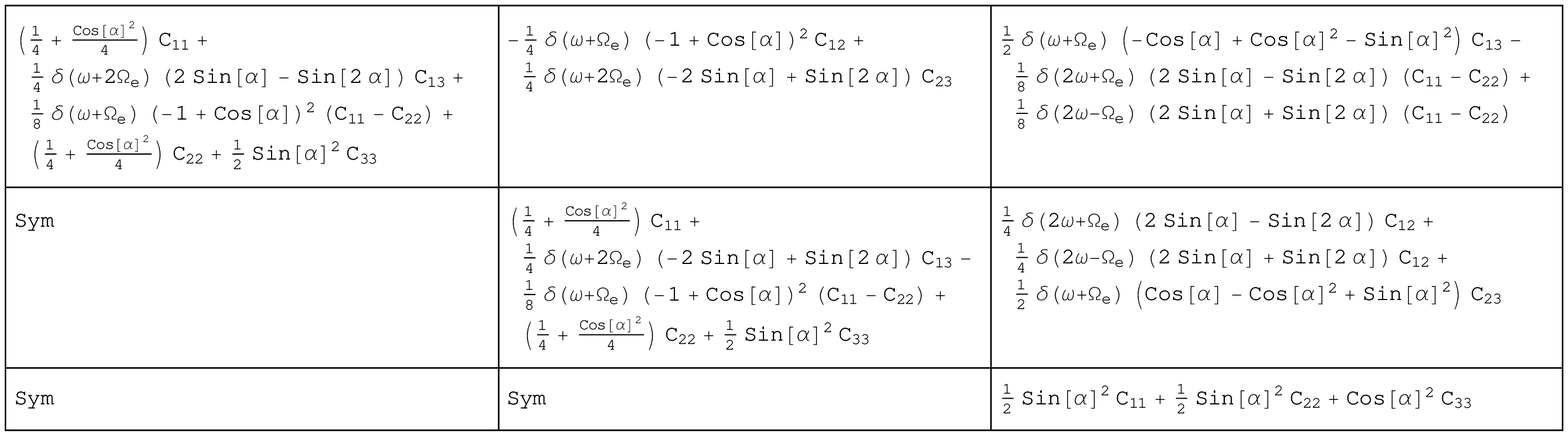}
	\caption[Like spins : Time average value of $\A(ij)$]{Like spins : Time average of the matrix $\A(ij)$. All time-averaged Hamiltonians for like spins presented in Section~IVA of the main article follow from this expression.}
	\label{fig:Alike Av an}
\end{figure}
\pagebreak
\section{Unlike spins}\label{sec:Results unlike an}

For unlike spins, $\gamma_i\neq\gamma_j$. The main article associates indices $i$ and $j$ with two different spin species. In this section, in order to make formulas better readable, we change the notations as follows:
\begin{equation}
	\alpha_i\equiv\alpha
	\label{eq:ABuni an}
\end{equation}
\begin{equation}
	\alpha_j\equiv\beta
	\label{eq:ABunj an}
\end{equation}
\begin{equation}
	\Ome[i]\equiv\Omega.
	\label{eq:OmUi an}
\end{equation}
\begin{equation}
	\Ome[j]\equiv\Lambda.
	\label{eq:OmUj an}
\end{equation}

In the present case, $\Pij(i)\neq \Q(j)$, and, therefore, $\A(ij)$ is not symmetric. The time-dependent matrix $\A(ij)$ is presented as a function of $\Cij(ij),\;\Pij(i)$ and $\Q(j)$ in Fig.~\ref{fig:Aunlike PQ an}. The explicit form of $\A(ij)$ as a function of $\alpha$, $\beta$ $\Om$, $\Omega$ and $\Lambda$ is given in Fig.~\ref{fig:Aunlike Om an}. The result of time-averaging for $\A(ij)$ is presented in Fig.~\ref{fig:Aunlike Av an}. 
\begin{figure}[h!]
	\centering
	\includegraphics[width=\textwidth]{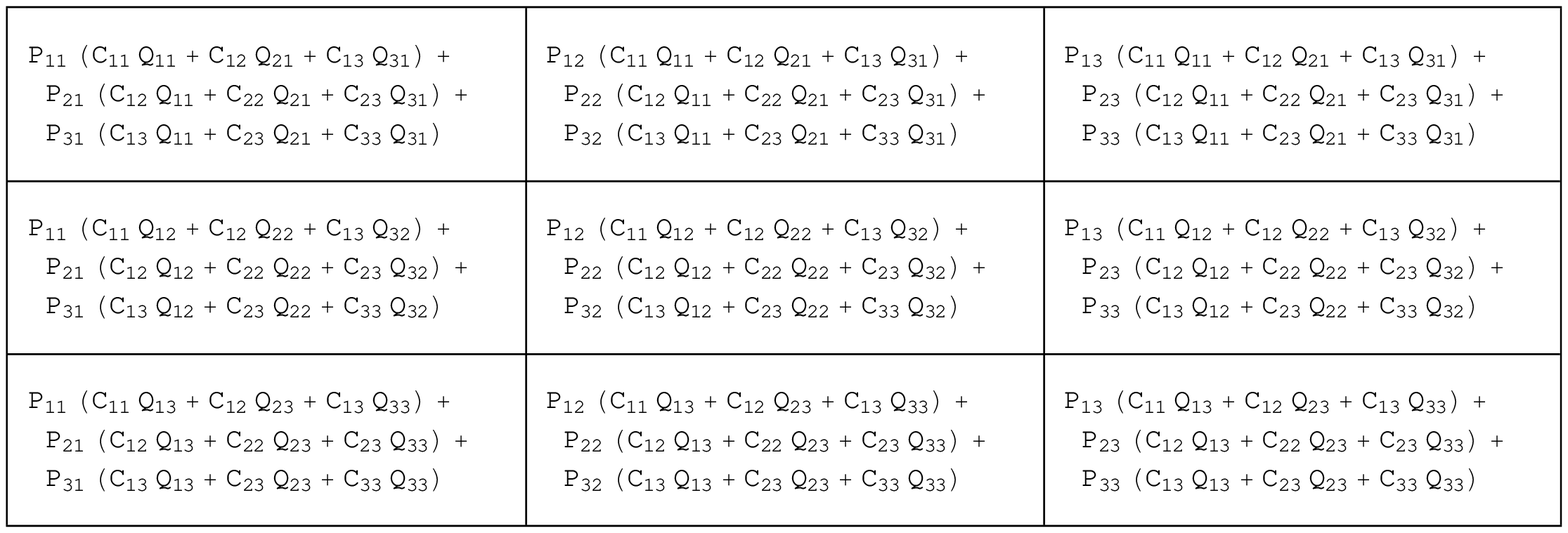}
	\caption{Unlike spins : The matrix $\A(ij)$ in compact form. Here $C_{\mu\nu} \equiv \Cij(ij)$, $P_{\mu\nu} \equiv \Pij(i)$ and $Q_{\mu\nu} \equiv \Q(j)$.}
	\label{fig:Aunlike PQ an}
\end{figure}
\vfill
\pagebreak
\begin{figure}[H]
	\centering
	\includegraphics[height=0.95\textheight,width=\textwidth]{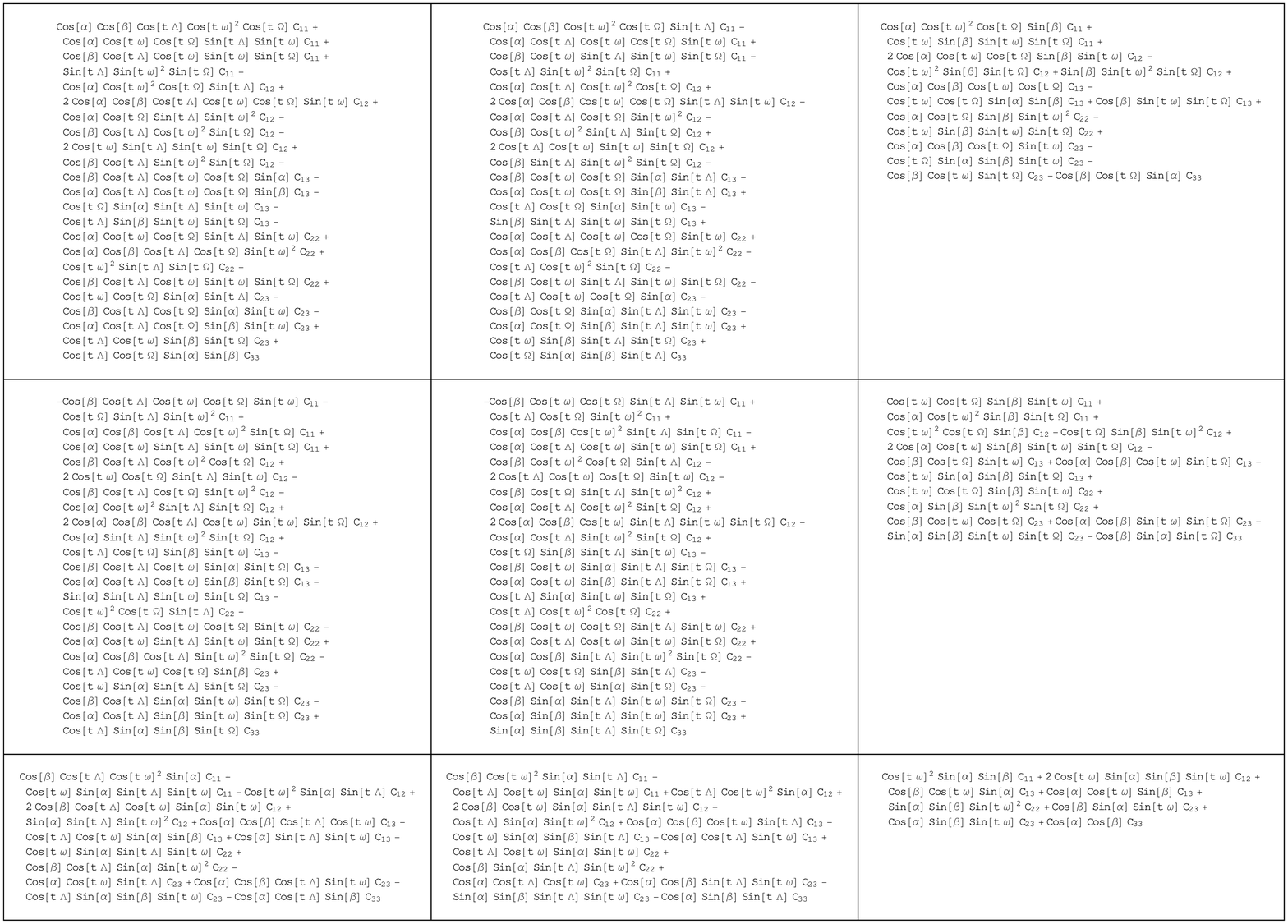}
	\caption{Unlike spins : Explicit form of the matrix $\A(ij)$}
	\label{fig:Aunlike Om an}
\end{figure}
\begin{figure}[H]
	\centering
	\includegraphics[height=0.45\textheight,width=1\textwidth]{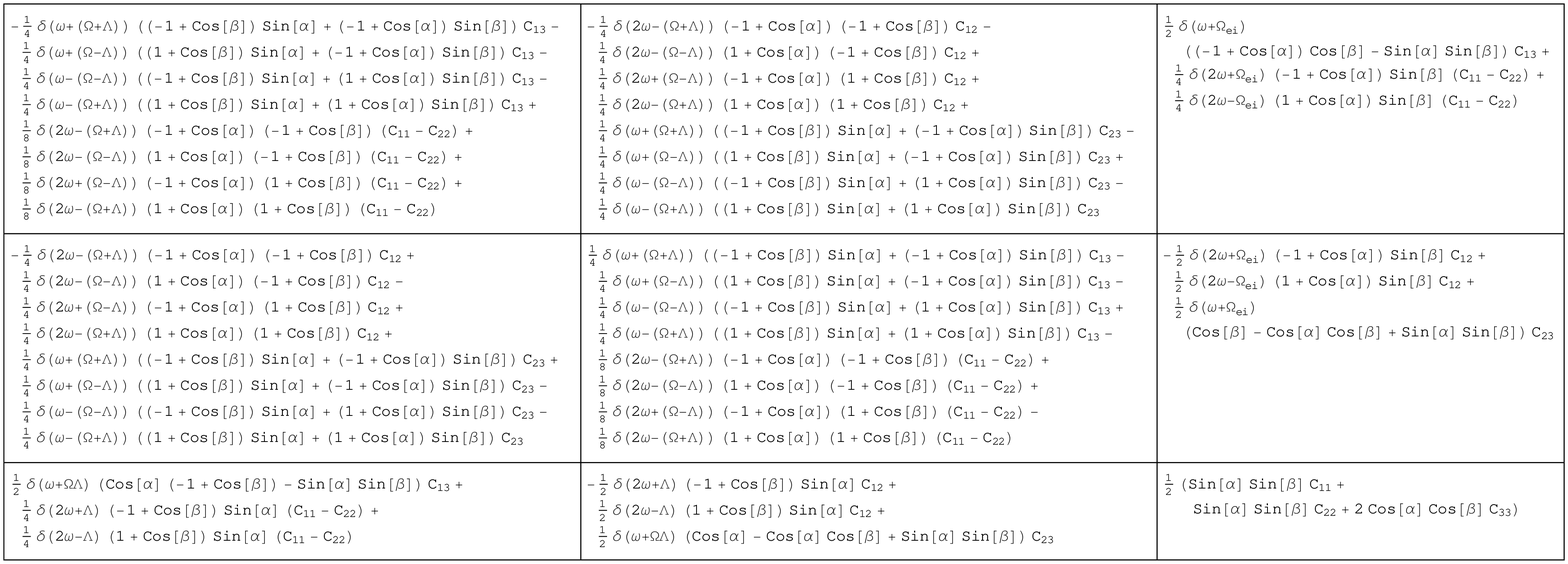}
	\caption[Unlike spins : time average value of $\A(ij)$]{Unlike spins : time average of the matrix $\A(ij)$. All time-averaged Hamiltonians for unlike spins presented in Section~IVB of the main article follow from this expression.}
	\label{fig:Aunlike Av an}
\end{figure}

\end{widetext}

\end{document}